\documentclass[journal]{IEEEtran}

\usepackage[T1]{fontenc}%
\usepackage[noadjust]{cite}

\usepackage{xcolor}
\usepackage{color,soul}
\usepackage[pdftex]{graphicx}
\graphicspath{{plots/pdf/},{drawings/}}
\usepackage{subfig}
\usepackage{amsmath}
\usepackage[cmintegrals]{newtxmath}
\usepackage[ruled,vlined]{algorithm2e}

\newtheorem{thm}{Theorem}
\newtheorem{lem}{Lemma}
\newtheorem{prop}{Proposition}

\begin{document}
\title{UAV Swarm Position Optimization for\\  High Capacity MIMO Backhaul}

\author{Samer~Hanna,~\IEEEmembership{Student~Member,~IEEE,}
        Enes~Krijestorac,~\IEEEmembership{Student~Member,~IEEE,}
        and~Danijela~Cabric,~\IEEEmembership{Fellow,~IEEE}%
\thanks{The authors are with the Electrical and Computer Engineering Department, University of California, Los Angeles, CA 90095, USA. e-mail: \mbox{samerhanna@ucla.edu}, enesk@g.ucla.edu, danijela@ee.ucla.edu.}%
\thanks{This work was supported in part by NSF  under  grant 1929874 and by the CONIX Research Center, one of six centers in JUMP, a Semiconductor Research Corporation (SRC) program sponsored by DARPA. }%
}

\maketitle

\begin{abstract}
A  swarm  of  cooperating UAVs  communicating  with a distant multiantenna ground station can leverage MIMO spatial multiplexing to scale the capacity.  Due to the line-of-sight propagation between the swarm and the ground station, the MIMO channel is highly correlated, leading to  limited  multiplexing gains. In this paper, we optimize the UAV positions to attain the maximum MIMO capacity given by the single user bound. An infinite  set of UAV placements that attains the capacity bound is first derived. Given an initial swarm placement, we formulate the problem of minimizing the distance traveled by the UAVs to reach a placement within the capacity maximizing set of positions. An offline centralized solution to the problem using block coordinate descent is developed assuming known initial positions of UAVs. We also propose an online distributed algorithm, where the UAVs iteratively adjust their positions to maximize the capacity.  Our proposed approaches are shown to significantly increase the capacity at the expense of a bounded translation from the initial UAV placements. This capacity increase persists when using a  massive MIMO ground station. Using numerical simulations, we show the robustness of our approaches in a Rician channel under UAV motion disturbances.  
\end{abstract}


\IEEEpeerreviewmaketitle

\section{Introduction}

\IEEEPARstart{D}{riven} by their low cost,  high mobility, and ease of deployment, 
unmanned aerial vehicles (UAVs) are used in many applications including delivery of goods,  surveillance, precision agriculture, and civil infrastructure inspection~\cite{shakhatreh_unmanned_2019}.  For wireless communications, UAVs have been proposed  as aerial basestations (BS)~\cite{alzenad_3-d_2018}, data aggregators~\cite{hattab_energy-efficient_2017}, and for many other applications~\cite{mozaffari_tutorial_2019}. The key advantage of UAV deployments over ground antennas is their mobility~\cite{mozaffari_tutorial_2019}. By changing the position of  UAVs, the wireless communications channel can be designed to be a  line-of-sight  channel for a given deployment~\cite{chen_efficient_2020}. However, unlike ground BSs which rely on wired communications for backhaul,  UAV  backhaul has to be  wireless, which makes it one of the challenges of UAV BSs~\cite{vinogradov_tutorial_2018-1}. In  scenarios requiring high throughput like open-air festivals~\mbox{\cite[Sec.~15]{fallgren_scenarios_2013}}, multiple UAVs need to be deployed  due to the lack of nearby  fixed infrastructure. In such multi-UAV deployments, the UAVs are  positioned close to the ground users, yielding a high SNR access link. In contrast, the backhaul link with a distant ground station has lower SNR and is shared among multiple UAVs thus creating a  bottleneck for communications.

Using a multiantenna ground station (GS)  makes the backhaul  air-to-ground channel between the swarm of UAVs and the GS  a MIMO channel.   By leveraging spatial multiplexing in this  channel,  the backhaul capacity can significantly be improved. However, the air-to-ground channel between a UAV swarm and the GS is typically dominated by line-of-sight (LOS) propagation~\cite{khuwaja_survey_2018}. Additionally, the  GS antenna array is typically implemented as a uniform linear or rectangular array  with limited dimensions. These factors can lead to highly correlated low-rank channels that can limit the MIMO capacity gains~\cite[Sec. 7.2.3]{tse_wireless_2005}. Optimizing the UAV swarm positions can reduce the channel correlation and improve capacity. But, this optimization for backhaul capacity should have a minimum impact on the  UAV  deployment application. 

The concept of positioning antennas to optimize the MIMO channel  was first proposed for LOS communications between fixed GSs. By optimizing  the spacing of  uniform antenna arrays, the maximum capacity given by the single user bound can be achieved~\cite{knopp_measurements_2007}. Once synchronized~\cite{yan_aeroconf_2019}, a cooperative UAV swarm can be viewed as a virtual antenna array. But unlike  GSs, which are limited in size and typically consist of uniformly spaced antenna elements, UAV swarms can achieve large apertures and are not bound to any geometry.  This flexibility  allows a swarm to maximize the backhaul capacity at the cost of moving the UAVs from their initial  positions~\cite{hanna_icassp_2019}. While it is possible to optimize for access and backhaul link simultaneously~\cite{fouda_interference_2019}, this limits the scope of the problem to  UAV BSs. Communications intensive UAV applications tolerating some displacement   are numerous and can include some video surveillance and remote sensing deployments~{\cite{shakhatreh_unmanned_2019}}.

In this paper, for a communications intensive application, we  optimize the placements of UAVs within a swarm to attain the maximum MIMO capacity  \textit{}over a backhaul link with a distant uniform rectangular GS.  Using the GS geometry and the LOS channel, we derive a set of UAV positions that attain the maximum capacity given by the single user bound.  Among the  capacity maximizing position set, UAV positions closer to  their initial placements pose the least disturbance to the deployment application and are desirable. Based on that, we formulate the problem of finding the positions within the capacity maximizing set that minimize the traveled distance from the initial positions. Two methods are considered to solve this problem; the first one is a centralized offline solution assuming a prior knowledge of the UAVs' initial positions. The second approach is a distributed online approach, where the UAVs iteratively adjust their positions to maximize the backhaul capacity.
 Our contributions are
\begin{itemize}
	\item We show that, for a uniform rectangular array ground station, the set of UAV placements  maximizing the MIMO capacity  is infinite. We derive a subset of placements within this set in the far-field  where the distance from the swarm to the ground station is much larger than the size of the swarm.
	\item Given the UAVs' initial placements, we formulate the problem of finding UAV positions within the capacity maximizing set that minimize the distance traveled.  A centralized suboptimal solution to this problem using block coordinate descent is developed and shown to require a bounded traveled distance  per UAV.
	\item An online distributed algorithm is proposed to maximize the capacity, requiring only sharing of channel estimates between neighboring UAVs. The conditions needed to guarantee its convergence  and an upper bound for the traveled distance are derived.
\end{itemize}

\section{Related Work}
Many of the existing works have focused on optimizing the UAV positions to improve only the access link~\mbox{\cite{alzenad_3-d_2018,lai_-demand_2019,krijestorac_uav_2019}}, thus implicitly assuming an ideal backhaul link. However, as the number of UAVs increase, wireless backhaul becomes challenging~{\cite{vinogradov_tutorial_2018-1}}. Some  works have proposed different approaches for UAV backhaul which we briefly discuss. We also discuss MIMO in UAV networks; some works has envisioned massive MIMO BS serving UAVs, others have proposed multiantenna UAV  BSs. Swarms of single antenna UAVs were also proposed communicating with either ground  users or   GSs.

\paragraph{Approaches for UAV Backhaul }
To address the challenges associated with UAV wireless backhaul, several approaches were proposed in the literature.  
 Some works have proposed using mmWave backhaul~\mbox{\cite{bertizzolo_mmbac_2019,tafintsev_aerial_2020}}, however, the high path loss at these frequencies makes them unsuitable for long links. Mechanical antenna steering was  proposed for UAVs backhaul at microwave frequencies~{\cite{pokorny_concept_2018}}, but  mechanical steering limits beamforming to one direction and is inherently slower. Integrated access and backhaul (IAB) links optimization  was proposed in~\mbox{\cite{youssef_backhaul-constrained_2019,fouda_uav-based_2018,fouda_interference_2019}}, where UAVs relay data using the same frequency bands in both links.  In \mbox{\cite{youssef_backhaul-constrained_2019}},  the locations, power allocation, and frequency assignment of a swarm of UAVs were optimized to reduce the transmit power in an IAB network using  a single antenna GS.  The  UAVs' frequency assignment  is chosen to minimize the interference between the backhaul and access links, which share the same bands.
  In~{\cite{fouda_uav-based_2018}}, using IAB,  exhaustive search is used to determine the UAV locations, precoder design, power allocations, which maximize the sum-rate to the ground users using a massive MIMO GS for backhaul. However, due to the prohibitive complexity of exhaustive search, only one UAV was considered. A less complex centralized solution to the same problem  using a fixed point method and particle swarm optimization was proposed in~{\cite{fouda_interference_2019}}. Results have shown that increasing the number of UAVs  increases the interference in the access link and reduces the network performance. In IAB networks,   the main challenge is to minimize the interference in the access link and between both access and backhaul links since the same frequencies are reused.    In this paper, our focus is on maximizing the swarm MIMO backhaul capacity for any application tolerating displacement from a given initial placement. In the case of UAV  BSs, the access link is assumed to use a different frequency band than the backhaul link. 

 \paragraph{UAVs Served by Massive MIMO BS}
In~\mbox{\cite{geraci_supporting_2018,garcia-rodriguez_essential_2019}}, massive MIMO cellular BSs  were proposed for cellular-connected UAVs and are shown to improve the data rates. Unlike our work, the UAVs are treated as user equipment with no control over their positions.
 Deep reinforcement learning was also proposed  for  navigation of a single UAV communicating with a massive MIMO BS in~{\cite{huang_deep_2020}}. The impact of having a swarm of UAVs on the  capacity was not considered.
 \paragraph{MIMO using multiantenna  UAVs}
  Using UAVs carrying  antenna arrays was  proposed in many works in the literature.
Some works have considered optimizing the positions and the beamforming vectors  to improve the ground users' SNR~{\cite{rupasinghe_uav_ang_2016}} or minimize the transmit power~{\cite{xu_robust_2018}}. The trajectory of  multiantenna UAVs serving ground users under an uncertain environment was optimized in~\mbox{\cite{xu_multiuser_2019}}.  However, for UAVs carrying an antenna array, the  UAV size and maximum payload for safe flight significantly constrains  the antenna array aperture compared to a swarm of single antenna UAVs, thus limiting the multiplexing gains with a distant GS.

\paragraph{ MIMO using UAV Swarms}
Several existing works have proposed   UAV swarms  leveraging MIMO. These works have either considered  the access link with ground users or the backhaul link with a GS.  For the access link, in~\cite{mozaffari_uav_array_2017}, the motion and beamforming weights of linearly arranged UAV swarm were optimized to serve users one at a time. To serve multiple  ground users simultaneously, UAVs were proposed  as  remote radio heads in a coordinated multipoint (CoMP) system and were optimized  to improve the capacity in~\cite{liu_comp_2018} and  physical layer security in~\cite{wang_uav_2019}.  Access link air to ground channel  is  different from the backhaul channel; in the former, the ground users are typically spread out, closer to the UAVs, and are more likely to get obstructed  unlike a distant  GS with a dominant LOS channel in the latter. 

To improve the capacity in LOS channels,  before the interest in UAV networks, the designs of traditional uniform antenna arrays   like linear and rectangular were optimized in~\mbox{\cite{bohagen_ula_2007,bohagen_ura_2007}}. Based on these designs, uniform geometries were proposed for UAV swarms communicating with GS in~\cite{su_maximum_2013,hanna_spawc_2019,pogue_IROS_2020}.  However, these rigid geometrical placements might conflict with positions required by application-driven deployments.  In~\cite{chandhar_massive_2018}, for a given UAV deployment,  the massive MIMO GS was optimized to maximize the ergodic LOS channel capacity, thus not benefiting from the UAVs mobility and requiring GS redesign per deployment. In~\cite{madhow_random_2013}, the authors proposed randomly placing the UAVs within a specified area for optimal MIMO capacity. Due to the randomness of this approach, the capacity improvements are probabilistic  and a large capacity increase requires having  more UAVs than GS antennas. 
 In~\cite{hanna_icassp_2019}, two iterative distributed algorithms were proposed to optimize the LOS MIMO  channel capacity of a UAV swarm; namely gradient descent and brute force, which are described later in this work.  However, no convergence proofs nor travel upper bounds were developed for the proposed algorithms. 
In this work, we leverage the UAV mobility to optimize the backhaul link capacity. Our proposed approaches  minimize the UAVs' displacements from given initial positions. Upper bounds on the distance traveled  and convergence proofs are derived for our proposed approaches.

\renewcommand{\b}[1]{\boldsymbol{\mathrm{#1}}}
\providecommand{\h}[1]{\ensuremath{\b{h}_{#1}}}

\newcommand{\C}[1]{\mathbb{C}^{#1}}
\newcommand{\R}[1]{\mathbb{R}^{#1}}
\newcommand{\Z}{\mathbb{Z}}
\newcommand{\U}{\mathcal{U}}
\newcommand{\mI}{\b{I}}
\newcommand{\E}[1]{\mathbb{E}\{#1\}}

\newcommand{\mGs}{\mathrm{G}}
\newcommand{\mUav}{\mathrm{U}}

\newcommand{\mNg}{M}
\newcommand{\mNgx}{M_x}
\newcommand{\mNgz}{M_z}
\newcommand{\mDgx}{d_x}
\newcommand{\mDgz}{d_z}
\newcommand{\mDux}{e_x}
\newcommand{\mDuz}{e_z}
\newcommand{\mNu}{N}
\newcommand{\mIg}{m}
\newcommand{\mIu}{n}

\newcommand{\mIx}{i}
\newcommand{\mIz}{j}

\newcommand{\mIux}{i_{\mIu}}
\newcommand{\mIuz}{j_{\mIu}}
\newcommand{\mIuxc}{i_{c}}
\newcommand{\mIuzc}{j_{c}}

\newcommand{\mIgx}{i_\mIg}
\newcommand{\mIgz}{j_\mIg}

\newcommand{\mIxl}{i_l}
\newcommand{\mIxk}{i_k}
\newcommand{\mIzl}{j_l}
\newcommand{\mIzk}{j_k}

\newcommand{\mPosU}[1]{\b{p}_{#1} }
\newcommand{\mPosUx}[1]{x_{#1} }
\newcommand{\mPosUy}[1]{y_{#1} }
\newcommand{\mPosUz}[1]{z_{#1} }

\newcommand{\mPosUyN}[1]{\epsilon_{#1} }

\newcommand{\mPosUxB}[1]{\overline{x}_{#1} }
\newcommand{\mPosUyB}[1]{\overline{y}_{#1} }
\newcommand{\mPosUzB}[1]{\overline{z}_{#1} }

\newcommand{\mPosG}[1]{\b{q}_{#1} }
\newcommand{\mPosUI}[1]{\overline{\b{p}}_{#1} }

\newcommand{\mPosRx}[1]{x'_{#1} }
\newcommand{\mPosRz}[1]{z'_{#1} }

\newcommand{\mPosRxd}[1]{x''_{#1} }
\newcommand{\mPosRzd}[1]{z''_{#1} }

\newcommand{\mPosRxB}[1]{\bar{x}'_{#1} }
\newcommand{\mPosRzB}[1]{\bar{z}'_{#1} }

\newcommand{\mPosRxD}[1]{\dot{x}_{#1} }
\newcommand{\mPosRzD}[1]{\dot{z}_{#1} }

\newcommand{\mP}{P}

\newcommand{\mH}{\b{H}}
\newcommand{\mHB}{\bar{\b{H}}}
\newcommand{\mhc}[1]{\b{h}_{#1} }
\newcommand{\mhe}[1]{h_{#1} }

\newcommand{\mFb}[1]{ \|#1\|_F }
\newcommand{\md}[1]{d_{#1} }

\newcommand{\mR}{R}

\newcommand{\mLos}{\text{LOS}}
\newcommand{\mNlos}{\text{NLOS}}

\newcommand{\mHLos}{\mH_{\mLos}}
\newcommand{\mHNlos}{\mH_{\mNlos}}
\newcommand{\mHLosB}{\overline{\mH}_{\mLos}}
\newcommand{\mHNlosB}{\overline{\mH}_{\mNlos}}

\newcommand{\mSnr}{\rho}
\newcommand{\mCmax}{C_{\text{max}}}
\newcommand{\mPset}{\mathcal{P}}
\newcommand{\mFset}{\mathcal{F}}
\newcommand{\mPmat}{\b{P}}
\newcommand{\mPmatD}{\widetilde{\b{P}}}

\section{System Model}
\begin{figure}[t!]
	\centering
	\includegraphics{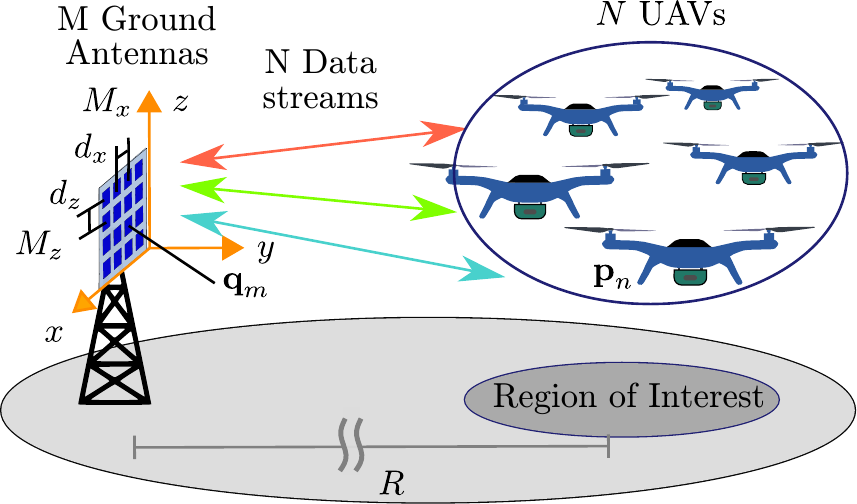}
	\caption{In the proposed system model, the UAV swarm is in the far-field of a uniform rectangular antenna array GS.}	
	\label{fig:system_model}
\end{figure}

A swarm of  $\mNu$ single antenna UAVs is communicating with a  multiantenna GS having $\mNg$  antennas in either uplink or downlink. Each UAV has its own data to transmit or to receive. To avoid  high inter-swarm communications overhead, the MIMO processing (precoding for downlink and combining for uplink) is done at the GS and each UAV sends or receives only its own stream. Since the maximum number of simultaneous streams possible is  $\min(\mNg,\mNu)$, at most $\mNu=\mNg$ UAVs can benefit from the MIMO gains. Hence, throughout this work, we assume that $\mNu\leq \mNg$. If $\mNu > \mNg$, time multiplexing or another technique has to be used, which is not considered in this work. 

The  channel between the swarm and  the ground antennas can be modeled as a Rician channel~\cite{khuwaja_survey_2018} and is denoted by $\mH\in \C{\mNg\times\mNu}$  defined as follows~\cite{bohagen_ula_2007}
\begin{equation}
\mH = \sqrt{\frac{K}{K+1}} \mHLos + \sqrt{\frac{1}{K+1}} \mHNlos
\end{equation}
where $K$ is the K-factor, $\mHLos$ is the  line-of-sight (LOS) component and $\mHNlos$ is the non-line-of-sight (NLOS) component. The elements of $\mHNlos$ are independently drawn from a zero-mean circularly symmetric complex Gaussian distribution with variance equal to  $\frac{\mFb{\mHLos}}{\mNg\mNu}$,where $\mFb{.}$ is the Frobenius norm. The normalized LOS channel $\mHLosB$ and   LOS channel accounting for path loss $\mHLos$   are given by 
\begin{equation}
\left[\mHLosB\right]_{\mIg,\mIu}= \exp \left( \frac{-j 2\pi \| \mPosU{\mIu} -\mPosG{\mIg} \|}{\lambda} \right)
\end{equation}
\begin{equation}
\left[\mHLos\right]_{\mIg,\mIu}= \frac{\lambda}{4 \pi \| \mPosU{\mIu} -\mPosG{\mIg} \|}\left[\mHLosB\right]_{\mIg,\mIu}
\end{equation}
where $\lambda$ is the wavelength and $\| \mPosU{\mIu} -\mPosG{\mIg} \|$ is the distance between the $\mIu$-th UAV and the $\mIg$-th GS antenna.  The element of the $\mIg$-th row, $\mIu$-th column of a matrix $\b{X}$ is denoted by $\left[\b{X}\right]_{\mIg,\mIu}$.
The $\mIu$-th UAV is located at position $\mPosU{\mIu}\in \R{3}$  defined as $\mPosU{\mIu} =[\mPosUx{\mIu},\mPosUy{\mIu}, \mPosUz{\mIu}]$. Similarly, the $\mIg$-th ground antenna is located at position $\mPosG{\mIg}\in \R{3}$.    The  matrix $\mPmat \in \R{3 \times \mNu}$ contains all the UAV positions such that  $\mPmat=[\mPosU{0}^T,\cdots,\mPosU{\mNu-1}^T]^T$ with $()^T$ denoting the transpose.

The GS is assumed to be arranged as a $\mNgx \times \mNgz$ uniform rectangular array where $\mNg=\mNgx\times \mNgz$. Without loss of generality, the GS is assumed to be placed in the x-z plane with $\mPosG{0}=[0,0,0]^T$  used as a coordinate reference. The spacing between the antennas in the x and z directions is given by $\mDgx$ and $\mDgz$ respectively. Hence,   $\mPosG{\mIg}=[\mIgx \mDgx,0,\mIgz \mDgz]^T$, where $\mIgx$ and $\mIgz$ are the antennas indices in x and z directions respectively and satisfy $\mIg =\mIgx \mNgz + \mIgz$.
The average separation along the y-axis between the UAVs and the GS is given by  $\mR=\frac{\sum_{\mIu=0}^{\mNu-1}\mPosUy{\mIu}}{\mNu}$. The system model is illustrated in Fig.~\ref{fig:system_model}.

We assume that the UAV swarm operates  within a bounded region in the far-field and that the GS is pointed toward the swarm such that $\mPmat \in \mFset$ where $\mFset$  is a position set defined as
\begin{multline}
	\mFset = \bigg\{ \mPmat \  | \ \mPosUy{\mIu} >> |\mPosUy{\mIu} -  \mPosUy{\mIg}|, \  \mPosUy{\mIu} >> |\mPosUx{\mIu}|, \ \mPosUy{\mIu} >> |\mPosUz{\mIu}|  , \\
	 \mPosUy{\mIu}>>\mNgx \mDgx,\ \mPosUy{\mIu}>>\mNgz \mDgz ,\ 
	  \mIu,\mIg \in \{0,\cdots,\mNu-1\}\bigg\} \label{eq:Fset_1}
\end{multline}
 In this work,  the swarm is always assumed to be within the set $\mFset$.
 Using these assumptions, the  magnitude of all the elements of the LOS channel matrix can be approximated to be constant and equal to $\frac{\lambda}{4\pi \mR}$ such that 
 \begin{equation}
 \label{eq:equal_mag}
 \mHLos \approx \frac{\lambda}{4\pi \mR} \mHLosB
 \end{equation}
 	
The single user bound defines the maximum achievable capacity and is given by~\cite{ngo_aspects_2014}
\begin{equation}
\begin{aligned}
C & = \log \det \left( \mI + \mSnr \mH^H \mH  \right)\\
  & \leq\sum_{\mIu=0}^{\mNu-1}\log \left( 1 + \mSnr \| \mhc{\mIu}^{[c]}\|^2 \right) = \mCmax \label{eq:cap_ub}
\end{aligned}
\end{equation}
where $\mSnr$ is the signal to noise ratio and $\mhc{\mIu}^{[c]}$ is the $\mIu$-th column of $\mH$ and $()^H$ denote the Hermitian transpose. The maximum capacity given by the single user bound $\mCmax$ can be attained when the columns of the channel matrix are mutually orthogonal~\cite{ngo_aspects_2014}.  When the bound is reached, the $\mNu$ different data streams do not interfere with each other.  Using the magnitude approximation,  the channel maximizing the capacity  has to realize
\begin{equation}\label{eq:orthogonal_mat}
	\E{\mH^H \mH} = \mNg \left(\frac{\lambda}{4\pi\mR}\right)^2 \mI
\end{equation}

where $\mI$ is the $\mNu\times\mNu$ identity matrix and  $\E{}$ denotes the expectation with respect to the channel NLOS component. Equation ({\ref{eq:orthogonal_mat}}) defines the condition on the channel matrix to attain the single-user bound capacity $\mCmax$. However, to formulate an optimization problem over the UAV positions, we need to relate the UAV positions with ({\ref{eq:orthogonal_mat}}). Our objective is to define a set of UAV positions that realize equation  ({\ref{eq:orthogonal_mat}}) in order to maximize the capacity.  Later, this set is used in the problem formulation.

\section{Set of Capacity Maximizing Positions}
\label{sec:opt_pos}
\begin{figure}[t!]
	\centering
	\includegraphics[scale=0.8]{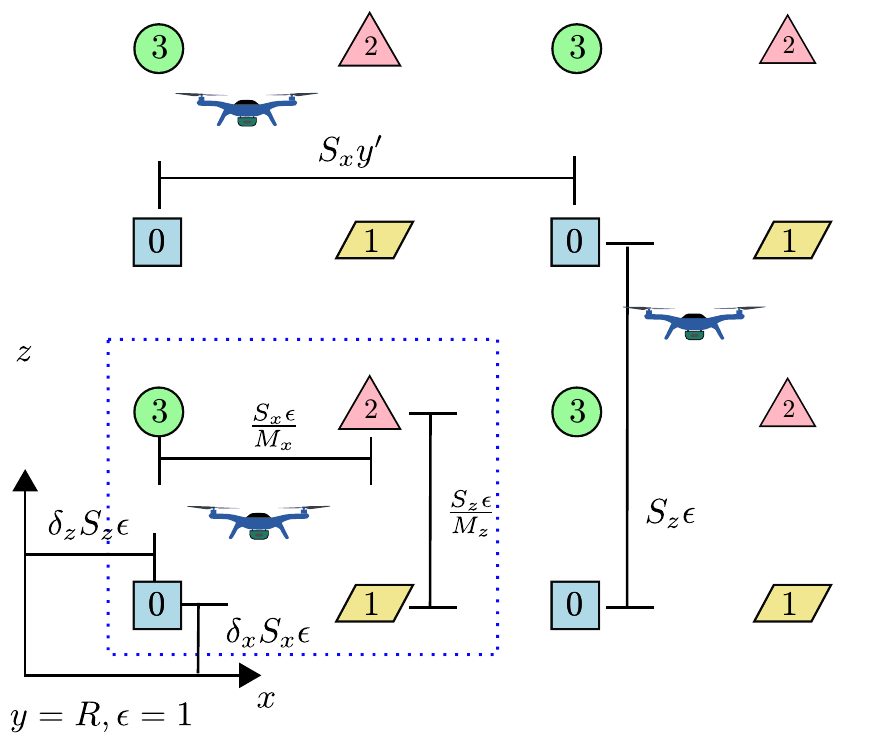}
	\caption{An illustration of the set  $\mPset$ for a swarm placed on the same plane at $y=R$ for a $2\times 2$ URA. The numbered colored shapes identify different positions. Each UAV needs to occupy a different position to maximize capacity.}	
	\label{fig:optimal_set}
\end{figure}
For an  $\mNg$ antenna GS, the maximum number of UAVs capable of using spatial multiplexing is $\mNg$. In this section, we aim to define the $\mNg$ positions for these UAVs that maximize the capacity.    If the number of UAVs, $\mNu$, is less than $\mNg$, the UAVs can be placed to occupy only $\mNu$ of these $\mNg$ positions. So without loss of generality, we consider  the matrix $\mH$ to be  a square matrix of size $\mNg\times \mNg$ and  (\ref{eq:orthogonal_mat}) can be rewritten in terms of the rows (instead of the columns) of $\mH$ as
\begin{equation} 
\label{eq:orthogonal_columns}
\E{\h{l}^{H} \h{k}} = \begin{cases}
\mNg \left(\frac{\lambda}{4\pi\mR}\right)^2  & l= k\\
0 &  l \neq k
\end{cases}  
\end{equation}
where $\h{k}$ is the $k$-th column of the transposed channel $\mH^T$.

Since $\mH$ is a Rician channel, due to the NLOS component $\mHNlos$, its elements are random variables. In our derivations, we consider the expected value of the columns inner product given by $\E{\h{l}^{H} \h{k}}$. Given our assumption that the elements $\mHNlos$ are independent and zero mean, $\E{\h{l}^{H} \h{k}} = \h{l}^{H[\text{LOS}]} \h{k}^{[\text{LOS}]}$ for $l\neq k $ where $\h{k}^{[\text{LOS}]}$ is the $k$-th column of $\mHLos^T$. Hence, using the LOS component in our derivations is equivalent to using the average over the Rician channel. To simplify the notations, we drop the expectation operator and we consider the channel matrix to be normalized such that 
$\mH=\mHLosB$.

We start by relating the right-hand side of   (\ref{eq:orthogonal_columns}) to the UAV positions as follows
\begin{align}
\h{l}^{H} \h{k} = \sum_{\mIu=0}^{\mNu-1} \exp \left( \frac{-j 2\pi }{\lambda} (\| \mPosU{\mIu} -\mPosG{l} \| - \| \mPosU{\mIu} -\mPosG{k} \|) \right)  \label{eq:strt}
\end{align}
To simplify the exponent of (\ref{eq:strt}),  we use our far-field assumption to approximate  the distance
\begin{equation}
\begin{aligned}
\|\mPosU{\mIu} - \mPosG{l} \| & =\sqrt{ (\mPosUx{\mIu}-\mIxl\mDgx)^2 + (\mPosUy{\mIu})^2+ (\mPosUz{\mIu}-\mIzl \mDgz)^2 }\\
&  \approx \mPosUy{\mIu}   \left(  1+ \frac{1}{2}\left(\frac{\mPosUx{\mIu}-\mIxl\mDgx}{\mPosUy{\mIu} }\right) +  \frac{1}{2}\left(\frac{\mPosUz{\mIu}-\mIzl \mDgz}{\mPosUy{\mIu} }\right)  \right) \label{eq:dist_rel}
\end{aligned}
\end{equation}
where $\mIxl$ and $ \mIzl$ are the  antenna indices along x and z respectively  and both satisfy  $l =\mIxl \mNgz + \mIzl$. The approximation uses the first-order Taylor approximation of the square root assuming that the UAVs are within $\mFset$. Hence, we can rewrite 
\begin{align}
\h{l}^{H} \h{k} & \approx  \sum_{\mIu=0}^{\mNu-1} \exp \left( 
\frac{-j 2\pi}{\mPosUy{\mIu} \lambda} \left( (-\mIxl+\mIxk) \mDgx \mPosUx{\mIu} + (-\mIzl+\mIzk) \mDgz \mPosUz{\mIu} \right)
 \right) 
 \label{eq:orthogonal_column_approx}
\end{align}
Note that from this result, we  can see that the  rate of change of the phase with respect to $\mPosUx{\mIu}$ and $\mPosUz{\mIu}$ is proportional to $1/\mPosUy{\mIu}$,  while the rate of change with respect to $\mPosUy{\mIu}$ is proportional to $1/\mPosUy{\mIu}^2$. Hence, to incur a given phase difference by moving the UAV in the y-direction requires  a much larger motion than by moving in the x or z directions. This observation will be used later when optimizing the UAV positions.

We start by describing one value of swarm positions $\mPmat$ that make the channel orthogonal in Lemma~{\ref{lem:grid_rect}}. After determining these positions, we investigate the changes in positions that retain orthogonality in Lemmas~{\ref{lem:grid_shifts}} and~{\ref{lem:grid_jumps}}. The set of positions maximizing capacity is obtained by combining these Lemmas in Theorem~{\ref{thm:pset}}.
\begin{lem}
	\label{lem:grid_rect}
	A uniform rectangular arrangement of UAVs within $\mFset$ having positions given by  $\mPosUx{\mIu} = \mIux \frac{\lambda \mPosUy{\mIu}}{\mNgx \mDgx}$ in the x-direction and $\mPosUz{\mIu} = \mIuz \frac{\lambda \mPosUy{\mIu}}{\mNgz \mDgz}$ in the z-direction realize channel  orthogonality, where $\mIux \in \{0,\cdots,\mNgx-1\}$ and $\mIuz \in \{0,\cdots,\mNgz-1\}$ such $\mIu = \mIux \mNgz + \mIuz$.
\end{lem}
\begin{IEEEproof}
	See Appendix~A.
\end{IEEEproof}
Lemma~\ref{lem:grid_rect} determines one value of swarm positions $\mPmat$, which makes the channel orthogonal. Now, we consider a few changes to positions that do not affect orthogonality.

\begin{lem}
	\label{lem:grid_shifts}
	Shifting the UAV swarm in the x-z plane such that each UAV is shifted  proportionally to its y separation from the GS does not affect channel orthogonality as long as the swarm remains within $\mFset$.
\end{lem}
\begin{IEEEproof}
	Let the set of UAV positions $\mPmat$ realize $\h{l}^{H} \h{k}=0$ for all $l\neq k$. Let $\h{l}'$ and  $\h{k}'$ be columns of the channel after shifting  UAV $\mIu$ by $\delta_x \mPosUy{\mIu}$ for all $\mIu$ in the x-direction and $\delta_z\mPosUy{\mIu}$ in the z-direction.
	\begin{equation}
	\begin{aligned}
	\h{l}^{'H} \h{k}'& =   \sum_{\mIu=0}^{\mNu-1} \exp \left( 
	\frac{-j 2\pi}{\mPosUy{\mIu} \lambda} \left( (-\mIxl+\mIxk) \mDgx \left(\mPosUx{\mIu} + 
	\delta_x\mPosUy{\mIu}\right) 
	\right.  \right.\\ &\left.  \left.   \ \ \ \ \ \ 
	+ (-\mIzl+\mIzk) \mDgz \left(\mPosUz{\mIu}+\delta_z\mPosUy{\mIu}\right) \right) 
	\right) \\
	& = \exp \left( \frac{ -j 2\pi}{\lambda}  ((-\mIxl+\mIxk) \delta_x + (-\mIzl+\mIzk) \delta_z) \right) \h{l}^{H} \h{k}\\
	& = 0
	\end{aligned}
	\end{equation}
	Hence, UAV shifts scaled with respect to their y coordinate do not affect the orthogonality of the MIMO channel.		
\end{IEEEproof}

\begin{lem}
	\label{lem:grid_jumps}
	Translation of individual  UAVs  such that UAV $\mIu$  is translated by an integer multiple of  $  \frac{\lambda \mPosUy{\mIu}}{d_x} $  in x-direction and/or   $ \frac{\lambda \mPosUy{\mIu}}{d_z} $ in z-direction does not affect the channel orthogonality  as long as the swarm remains within $\mFset$.
\end{lem}
\begin{IEEEproof}
	Let the   channel have columns $\h{l}$ and $\h{k}$ for some GS antennas $l$ and $k$. Let all UAVs be translated independent of each other, such that UAV $\mIu$ is shifted by  $ f_{\mIu} \frac{\lambda \mPosUy{\mIu}}{\mDgx}$ and $ g_{\mIu} \frac{\lambda \mPosUy{\mIu}}{\mDgz}$  in the x and z-direction respectively  for some UAV specific integers $f_{\mIu}$ and $g_{\mIu}$.  Let $\h{l}'$ and  $\h{k}'$ be columns of the channel after the shifting. Calculating the inner product of these columns, we get
	\begin{equation}
	\begin{aligned}
	\h{l}^{'H} \h{k}'& =   \sum_{\mIu=0}^{\mNu-1} \exp \left( 
	\frac{-j 2\pi}{\mPosUy{\mIu} \lambda} \left( (-\mIxl+\mIxk) \mDgx \left(\mPosUx{\mIu} + 
	f_{\mIu} \frac{\lambda \mPosUy{\mIu}}{\mDgx} \right) 
	\right.  \right.\\ &\left.  \left.   \ \ \ \ \ \ 
	+ (-\mIzl+\mIzk) \mDgz \left(\mPosUz{\mIu} + g_{\mIu} \frac{\lambda \mPosUy{\mIu}}{\mDgz}\right) \right) 
	\right) \\
	& = \exp \left( -j 2\pi ((-\mIxl+\mIxk) f_{\mIu} + (-\mIzl+\mIzk) g_{\mIu}) \right) \h{l}^{H} \h{k}\\
	& = \h{l}^{H} \h{k}
	\end{aligned}
	\end{equation}
\end{IEEEproof}
In addition to scaled swarm translations from Lemmas~\ref{lem:grid_shifts} and individual UAV jumps from Lemma~\ref{lem:grid_jumps}, it easy to see that permuting the positions of the swarm  does not affect orthogonality.  We define a set of positions orthogonalizing the channel by combining the three previous lemmas as follows.

\begin{thm}
	\label{thm:pset}
	Given a URA GS with $\mNg$ antennas, the  set $\mPset \subset \mFset$  is a set containing  placements, $\mPmat \in \R{3\times \mNg}$, of $\mNg$ UAVs,  which realize  the channel orthogonality condition given by ($\ref{eq:orthogonal_columns}$).  The set $\mPset$ can be described given the environment constants $S_x=\frac{\lambda \mR}{\mDgx}$ and $S_z=\frac{\lambda \mR}{\mDgz}$ as follows 
	\begin{multline}
	\label{eq:optimal_set}
		\mPset = \{  \mPmat : \mPmat =\b{T}_\Pi \mPmatD ,  \ \mPmatD \in \mFset,
		\ifdefined \singleCol\else	\\ \ \	\fi
		\mPosUx{\mIu} = \left[\mPmatD\right]_{0,\mIu}  , \mPosUyN{\mIu} = \frac{\left[\mPmatD\right]_{1,\mIu}}{\mR}  ,\mPosUz{\mIu} = \left[\mPmatD\right]_{2,\mIu}   \\
		\mPosUx{\mIu} = \mIux \frac{S_x \mPosUyN{\mIu}}{\mNgx} + f_{\mIu} S_x \mPosUyN{\mIu} + \delta_x S_x   \mPosUyN{\mIu},
		\ifdefined \singleCol\else	\\ \ \	\fi
		 \mPosUz{\mIu} = \mIuz \frac{S_z \mPosUyN{\mIu}}{\mNgz} + g_{\mIu}  S_z \mPosUyN{\mIu} + \delta_z   S_z  \mPosUyN{\mIu}\ , 
		 \ifdefined \singleCol\else	\\ \ \	\fi
		 \mIu = \mIux \mNgz + \mIuz , \\
		 \mIux \in \{0,\cdots,\mNgx-1 \},  \mIuz \in \{0,\cdots,\mNgz-1 \} , 
		 \ifdefined \singleCol\else	\\ \ \	\fi
		   \b{T}_\Pi \in \Pi^{\mNg} ,
		   \delta_x,\delta_z \in \mathbb{R}, f_{\mIu},g_{\mIu} \in \Z \forall \mIu\}
	\end{multline}
	where $ \Pi^{\mNg}$ is the set of all  $\mNg \times \mNg$ permutation matrices.
\end{thm}
\begin{IEEEproof}
	This can be proved by the  application of Lemmas \ref{lem:grid_shifts},  \ref{lem:grid_jumps}, and applying permutations on the results obtained by Lemma \ref{lem:grid_rect}. We renamed $\mPosUy{\mIu}=\mPosUyN{\mIu} \mR$ and  the shifts  from Lemma \ref{lem:grid_shifts} to  $\delta_x S_x   \mPosUyN{\mIu}$ and $\delta_z   S_z  \mPosUyN{\mIu}$ for convenience of notation.
\end{IEEEproof}

If the swarm positions are within $\mPset$ as defined in Theorem~{\ref{thm:pset}}, the orthogonality condition ({\ref{eq:orthogonal_mat}}) is satisfied and $\mCmax$ is attained. By using this set, an optimization problem over UAV positions can be formulated. We show an example of the set $\mPset$ in Fig.~\ref{fig:optimal_set} for a simple scenario having $\mNg=4$ with $\mNgx=\mNgz=2$ and $\mNu=3$ UAVs on the same x-z plane. 
The positions defined according to Lemma \ref{lem:grid_rect} are labeled 0 to 3 in different colored shapes inside the blue dotted square. According to Lemma \ref{lem:grid_jumps}, The $S_x\epsilon$ and the $S_z\epsilon$  jumps along x and z respectively are also within  $\mPset$.  This entire grid can be shifted by $\delta_x$ and $\delta_z$ according to Lemma~\ref{lem:grid_shifts}. Orthogonality is attained by  assigning the 3 UAVs to any of the 4 positions.  After defining the set of UAV positions, $\mPset$,  orthogonalizing the channel and attaining $\mCmax$, we discuss the problem formulation.

\section{Placement Optimization Problem Formulation}
 There are several ways to mathematically formulate the problem of  optimizing swarm positions to attain the maximum capacity. The most intuitive way is to optimize over the positions with the capacity as the objective. However, since  any swarm positions $\mPmat \in \mPset$ can achieve $\mCmax$ and no unique solution exists, using this formulation, the obtained positions can be  far from the UAVs' initial positions and hence would cause unnecessary disturbance to the deployment application. Given that the considered deployment application prioritizes communications with no hard constraints on UAV displacement, defining a constraint on the distance traveled by the UAVs is not straightforward; if the constraint is too tight, a suboptimal capacity below $\mCmax$ will be achieved, if the constraint is too loose, the solution might lead to unnecessary travel by the UAVs.  The minimal distance  to attain $\mCmax$   differs from one deployment environment to the other and hence can not be used as a constraint.  Instead, we make minimizing the distance traveled  our objective. To guarantee that the maximum capacity is attained, we constrain the optimized UAV positions to be within the set $\mPset$, which attains $\mCmax$.  This formulation attains  the maximum capacity with the least  traveled distance.
 
 A mathematical formulation of the problem is as follows; Given $\mNu$ UAVs with initial positions  $\{\mPosUI{0},\mPosUI{1},\cdots,\mPosUI{\mNu-1}\}$, where $\mPosUI{\mIu}=[\mPosUxB{\mIu},\mPosUyB{\mIu},\mPosUzB{\mIu}]^T$. These initial positions are assumed to be determined by the deployment application. Our objective is to  find  the nearest UAV positions which belong to   $\mPset$.  This problem can be formulated as 

\begin{align}
& \underset{ \{\mPosU{\mIg}\},\{b_{\mIg,\mIu} \} }{\text{minimize}} & & \sum_{\mIg=1}^{\mNg} \sum_{\mIu=1}^{\mNu} b_{\mIg,\mIu} \|\mPosUI{\mIu}-\mPosU{\mIg}\| \label{prob:main}\\
& \text{subject to} & & [\mPosU{0},\cdots,\mPosU{\mNg-1}]\in \mPset \nonumber  \\
&  & & \sum_{\mIu=0}^{\mNu} b_{\mIg,\mIu} \leq 1 \ \ \ \  \forall \mIg, \nonumber  \ \
&  & & \sum_{\mIg=0}^{\mNg} b_{\mIg,\mIu} = 1 \ \ \ \  \forall  \mIu   \nonumber  \\
& & & b_{\mIg,\mIu} \in \{0,1\} \ \ \ \forall \mIg,\mIu  \nonumber  
\end{align}
The binary variable $b_{\mIg,\mIu}$  is used to assign each of the $\mNu$ UAVs (indexed using $\mIu$) to one of the $\mNg$ positions within $\mPset$ (indexed using $\mIg$). This problem formulation does not make any assumptions about whether the UAVs are transmitters or receivers and does not make any assumptions about the transmitter and receiver processing. Later in Section~{\ref{sec:linear_scenario}}, we consider an uplink scenario and derive the optimal linear precoders and combiners.
 
Using the definition of  $\mPset$ from Theorem \ref{thm:pset} in (\ref{prob:main}), the problem can be rewritten as
\begin{align}
& \underset{\substack{\{\mPosUx{\mIg}\}, \{\mPosUy{\mIg}\}, \{\mPosUz{\mIg}\},\\ \delta_x,\delta_z, \{\mPosUyN{\mIu}\}, \\ \label{prob:main_det}  \{f_{\mIg}\}, \{g_{\mIg}\}, \{b_{\mIg,\mIu}\}  }
   }{\text{minimize}}   \\  & & & \hspace{-15mm} \sum_{\mIg=0}^{\mNg-1} \sum_{\mIu=0}^{\mNu-1} b_{\mIg,\mIu} \sqrt{(\mPosUx{\mIg}-\mPosUxB{\mIu})^2 +(\mPosUy{\mIg}-\mPosUyB{\mIu})^2+(\mPosUz{\mIg}-\mPosUzB{\mIu})^2}  \nonumber \\
& \text{subject to}  & 
  & \mPosUx{\mIg} = \mIgx \frac{S_x \mPosUyN{\mIg}}{ \mNgx} + f_{\mIg}  S_x \mPosUyN{\mIg} + \delta_x  S_x  \mPosUyN{\mIg}, \ \ \ \forall \mIg  \nonumber \\ 
&  & &  \mPosUz{\mIg} = \mIgz\frac{S_z\mPosUyN{\mIg}}{\mNgz } + g_{\mIg} S_z \mPosUyN{\mIg} +  \delta_z S_z \mPosUyN{\mIg}, \ \ \ \forall \mIg   \nonumber\\
&  & &  f_{\mIu},g_{\mIu} \in \Z \ \ \ \forall \mIu   \nonumber\\
&  & & \sum_{\mIu=0}^{\mNu} b_{\mIg,\mIu} \leq 1 \ \ \ \  \forall \mIg , \ \
 \sum_{\mIg=0}^{\mNg} b_{\mIg,\mIu} = 1 \ \ \ \  \forall  \mIu   \nonumber \\
&  & & b_{\mIg,\mIu} \in \{0,1\} \ \ \ \forall \mIg,\mIu \nonumber\\ 
& & & -\frac{1}{2}\leq \delta_x   \leq  \frac{1}{2} ,  -\frac{1}{2} \leq \delta_z   \leq  \frac{1}{2}   \nonumber\\
& & & \mPosUy{\mIg} = \mR \mPosUyN{\mIg}  \nonumber
 \end{align}
where $\mIgx\in \{0,\cdots,\mNgx\}$ and  $\mIgz \in \{0,\cdots,\mNgz\}$ and both  satisfy  $\mIg = \mIgx \mNgz + \mIgz$.  In the current form, this problem is a non-convex mixed-integer problem that is not tractable. 

To solve this problem, we consider both an offline centralized solution in Section~{\ref{sec:cent}} and an online distributed algorithm in Section~{\ref{sec:FF}}. The centralized solution assumes the initial positions are known apriori and aims to relax and solve~(\ref{prob:main_det}). In the case where the UAVs are already deployed without  prior knowledge of their placements, a distributed online algorithm   where the UAVs iteratively improve their positions is also proposed.

\section{Centralized Offline Solution}
\label{sec:cent}
For the centralized solution, we start by relaxing problem ({\ref{prob:main_det}}) to make it more tractable prior to deriving its solution. An upper bound for the distance traveled and the time complexity are also discussed.
\subsection{Problem Relaxation}
We start by eliminating the y-translation variable.
\paragraph{Eliminating y-translation} As discussed previously  $\mPosUy{\mIg}$ has a small effect on the phase unlike a change in $\mPosUx{\mIg}$ and  $\mPosUz{\mIg}$. A UAV has to travel a much larger distance along the y direction compared to the x or z direction to incur a phase change.
So to simplify, we relax the problem by not optimizing over the y-translation, i.e, setting $\mPosUy{\mIg}=\mPosUyB{\mIg}$  for all UAVs. Hence, we only optimize  over $\mPosUx{\mIg}$ and $\mPosUz{\mIg}$. Given this simplification, the y term in the objective is equal to zero and $\mPosUyN{\mIg}$ becomes a constant for all $m$. The problem can be reformulated as
\begin{align}
& \underset{\substack{\{\mPosRx{\mIg,\mIu}\},  \{\mPosRz{\mIg,\mIu}\},\\ \{f_{\mIu}\}, \{g_{\mIu}\}\\ \{b_{\mIg,\mIu}\} \\ \delta_x,\delta_z}
}{\text{minimize}} & & \sum_{\mIg=0}^{\mNg-1}  \sum_{\mIu=0}^{\mNu-1} b_{\mIg,\mIu} \sqrt{(\mPosRx{\mIg,\mIu})^2 +(\mPosRz{\mIg,\mIu})^2} \label{eq:centr_prob1} \\
& \text{subject to}  & 
& \mPosRx{\mIg,\mIu} =  \mIgx \frac{S_x \mPosUyN{\mIu}}{ \mNgx}   + f_{\mIg} S_x  \mPosUyN{\mIu} + \delta_x S_x  \mPosUyN{\mIu} -\mPosUxB{\mIu}  ,  \forall \mIu, \mIg \label{eq:const_x_def}\\ 
&  & &  \mPosRz{\mIg,\mIu} = \mIgz \frac{S_z \mPosUyN{\mIu}}{ \mNgz}  + g_{\mIg}S_z \mPosUyN{\mIu} +  \delta_z S_x  \mPosUyN{\mIu} -\mPosUzB{\mIu}  ,   \forall \mIu, \mIg \label{eq:const_z_def} \\
&  & &  f_{\mIu},g_{\mIu} \in \Z \ \ \ \forall \mIu \label{eq:const_fg_def}\\
&  & & b_{\mIg,\mIu} \in \{0, 1\} \ \ \  \forall \mIu, \mIg \\
&  & & \sum_{\mIu=0}^{\mNu} b_{\mIg,\mIu} \leq 1 \ \ \ \  \forall \mIg , \ \
 \sum_{\mIg=0}^{\mNg} b_{\mIg,\mIu} = 1 \ \ \ \  \forall  \mIu  \\
& & & -\frac{1}{2} \leq \delta_x   \leq  \frac{1}{2}  , \ \
 -\frac{1}{2} \leq \delta_z   \leq  \frac{1}{2}  \label{eq:centr_prob1_delta}
\end{align}
The value of $\mPosRx{\mIg,\mIu}$ is the x-translation difference between the initial position of  UAV $\mIu$ and the optimal position $\mIg$ and  $\mPosRz{\mIg,\mIu}$ is similarly defined for the z-translation. The integers $f_{\mIg}$ and $g_{\mIg}$ define multiple possible solutions, however, we know that the optimal one is closer to the starting positions. We use this intuition to narrow the solution space.

\paragraph{Narrowing Solution Space} According to Lemma~\ref{lem:grid_jumps}, any integer value of  $f_{\mIg}$ and $g_{\mIg}$ can achieve orthogonality. However, the values of these variables  that minimize the translation are expected to be the ones closest to the starting positions of the UAVs. To simplify the problem using this intuition,  we start by rewriting the initial positions of the UAVs $\mPosUxB{\mIu}$ and $\mPosUzB{\mIu}$ as a function of our environment constants as follows $\mPosUxB{\mIu} = c'_{\mIu} S_x \mPosUyN{\mIu} $ where $c'_{\mIu}$ is the constant satisfying this relation.  By substituting in~(\ref{eq:const_x_def}), we get
\begin{equation}
\begin{aligned}
 \frac{\mIgx}{ \mNgx} S_x \mPosUyN{\mIu}  + f_{\mIg} S_x \mPosUyN{\mIu}  -\mPosUxB{\mIu} &=   S_x \mPosUyN{\mIu} \left( f_\mIg + \frac{\mIgx}{ \mNgx}    -c'_\mIu  \right)\\
 &=   S_x \mPosUyN{\mIu} \left( f_{\mIg} - (f'_{\mIu} + r'_{\mIu}) \right)\\
&=\tilde{x}_{\mIg,\mIu} + f_{\mIg,\mIu} S_x \mPosUyN{\mIu}
\end{aligned}
\end{equation}
where  $f'_{\mIu} = \left\lfloor \frac{\mIgx}{ \mNgx}    -c'_\mIu  \right\rfloor$ is an integer obtained by the floor operation and 
$r'_{\mIu} =   \left(\frac{\mIgx}{ \mNgx}    -c'_\mIu \right)- f'_{\mIu} $ has a magnitude smaller than one.
The distance  $\tilde{x}_{\mIg,\mIu}$ is defined as  $\tilde{x}_{\mIg,\mIu} = S_x \mPosUyN{\mIu}  r'_{\mIu} $ and satisfies
 $0 \leq \tilde{x}_\mIu < S_x \mPosUyN{\mIu} $. We define $f_{\mIg,\mIu} = f_\mIg -f'_\mIu$, which redefines the integer translations to use the initial positions of the UAV  $\mIu$ as a starting point. 
Hence, we can rewrite~(\ref{eq:const_x_def}) as 
\begin{equation}
\label{eq:norm_x_trans}
 \mPosRx{\mIg,\mIu} = \tilde{x}_{\mIg,\mIu} + f_{\mIg,\mIu} S_x  \mPosUyN{\mIu} + \delta_x S_x \mPosUyN{\mIu}
\end{equation}
similarly for the z-direction, we get
\begin{equation}
\label{eq:norm_z_trans}
\mPosRz{\mIg,\mIu} = \tilde{z}_{\mIg,\mIu} + g_{\mIg,\mIu} S_z  \mPosUyN{\mIu} + \delta_z S_z \mPosUyN{\mIu}
\end{equation}
\begin{prop}
	\label{prop:bounded_ints}
	The  value of $f_{\mIg,\mIu}$ and  $g_{\mIg,\mIu}$ that minimizes~(\ref{eq:centr_prob1}) is within the set $\{-1,0\}$ and is given by
		 \begin{equation}
		 \begin{aligned}
		 \hat{f}_{\mIg,\mIu}  
		 &=\begin{cases}
		 0 &  -\frac{1}{2}   S_x \mPosUyN{\mIu} \leq \tilde{x}_{\mIg,\mIu} + \delta_x S_x \mPosUyN{\mIu} <  \frac{1}{2} S_x \mPosUyN{\mIu} \\
		 -1 &  \frac{1}{2}   S_x \mPosUyN{\mIu} \leq \tilde{x}_{\mIg,\mIu} + \delta_x S_x \mPosUyN{\mIu} \leq  \frac{3}{2} S_x \mPosUyN{\mIu}  \\
		 \end{cases}\label{eq:fg_min}
		 \end{aligned}
		 \end{equation}
\end{prop}
\begin{IEEEproof}
	See Appendix~B. 
\end{IEEEproof}
 A similar result can be proved for $g_{\mIg,\mIu}$. Hence, among all the values of integer translations from the initial UAV positions,  $f_{\mIg,\mIu}$ and $g_{\mIg,\mIu}$,   we only need to consider the values of the nearest  translations from the UAV's initial locations.
\paragraph{The relaxed problem}
The problem is thus simplified  to
\begin{align}
& \underset{\substack{\{\mPosRx{\mIu}\},  \{\mPosRz{\mIu}\},\\ \{f_{\mIg,\mIu}\}, \{g_{\mIg,\mIu}\}\\ \{b_{\mIg,\mIu}\} \\ \delta_x,\delta_z}
}{\text{minimize}} & &   \sum_{\mIg=0}^{\mNg-1}  \sum_{\mIu=0}^{\mNu-1} b_{\mIg,\mIu} \sqrt{(\mPosRx{\mIg,\mIu})^2 +(\mPosRz{\mIg,\mIu})^2} \label{prob:centralized_local} \\
& \text{subject to}  & 
& \mPosRx{\mIg,\mIu} = \tilde{x}_{\mIg,\mIu} + f_{\mIg,\mIu} S_x \mPosUyN{\mIu}  + \delta_x  S_x \mPosUyN{\mIu} , \ \ \ \forall \mIu, \mIg \nonumber \\ 
&  & &  \mPosRz{\mIg,\mIu} =\tilde{z}_{\mIg,\mIu} + g_{\mIg,\mIu} S_z \mPosUyN{\mIu} + \delta_z  S_z \mPosUyN{\mIu}, \ \ \ \forall \mIu, \mIg \nonumber \\
&  & &  f_{\mIg,\mIu},g_{\mIg,\mIu} \in \{-1,0\} \ \ \ \forall \mIu \nonumber\\
&  & & b_{\mIg,\mIu} \in \{0, 1\} \ \ \  \forall \mIu, \mIg \nonumber\\
&  & & \sum_{\mIu=0}^{\mNu} b_{\mIg,\mIu} \leq 1 \ \ \ \  \forall \mIg , \ \ 
 \sum_{\mIg=0}^{\mNg} b_{\mIg,\mIu} = 1 \ \ \ \  \forall  \mIu  \nonumber\\
& & & -\frac{1}{2} \leq \delta_x   \leq  \frac{1}{2} , \ \ 
 -\frac{1}{2} \leq \delta_z   \leq  \frac{1}{2} \nonumber
\end{align}

 \subsection{Problem Solution}
The problem (\ref{prob:centralized_local}) still remains a non-convex mixed-integer problem. The difficulty in solving the problem is because the variables $\delta_x$ and $\delta_z$ are common to the entire swarm. We show that by for a given value of some variables the problem becomes tractable and we use that fact to solve the problem.
\paragraph{Solution  given $\delta_x$ and $\delta_z$} For a given value of $\delta_x$ and $\delta_z$,  the problem becomes tractable and it can be solved as follows: first, we minimize over $f_{\mIg,\mIu}$ and $g_{\mIg,\mIu}$ using (\ref{eq:fg_min}) since $\delta_x$ and $\delta_z$ are given. Once these values have been calculated, the square root term in the objective becomes a constant. 
What remains is to solve for  $b_{\mIg,\mIu}$, which becomes the following integer linear program
\begin{align}
\label{prob:centralized_local_assignment}
& \underset{ \{b_{\mIg,\mIu}\}}
{\text{minimize}} & & \sum_{\mIg=0}^{\mNg-1}  \sum_{\mIu=0}^{\mNu-1} b_{\mIg,\mIu} \sqrt{(\mPosRx{\mIg,\mIu})^2 +(\mPosRz{\mIg,\mIu})^2} \\
& \text{subject to} & & b_{\mIg,\mIu} \in \{0, 1\} \ \ \  \forall \mIu, \mIg \nonumber\\
&  & & \sum_{\mIu=0}^{\mNu} b_{\mIg,\mIu} = 1 \ \ \ \  \forall \mIg , \ \ 
\sum_{\mIg=0}^{\mNg} b_{\mIg,\mIu} \leq 1 \ \ \ \  \forall  \mIu  \nonumber
\end{align}
 This integer program can be shown to be equivalent to its real relaxation. This problem is indeed an assignment problem that can be solved in polynomial time using the Hungarian algorithm~{\cite{kuhn_hungarian_1955}}.
 \paragraph{Solution  given an assignment}
Again considering  (\ref{prob:centralized_local}), the challenge in solving for $\delta_x$ and $\delta_z$ is that they are multiplied by integer variables $b_{\mIg,\mIu}$. Given an assignment defining the values of $b_{\mIg,\mIu}$, the problem (\ref{prob:centralized_local})  becomes the following convex problem 
 \begin{align}
 \label{prob:centralized_local_delta}
 & \underset{\substack{\{\mPosRx{\mIu}\},  \{\mPosRz{\mIu}\},\\ \delta_x,\delta_z}}{\text{minimize}} & & \sum_{\mIg=0}^{\mNg-1}  \sum_{\mIu=0}^{\mNu-1} b_{\mIg,\mIu} \sqrt{(\mPosRx{\mIg,\mIu})^2 +(\mPosRz{\mIg,\mIu})^2} \\
 & \text{subject to}  & 
 & \mPosRx{\mIg,\mIu} = \tilde{x}_{\mIg,\mIu} + \hat{f}_{\mIg,\mIu} S_x \mPosUyN{\mIu}  + \delta_x  S_x \mPosUyN{\mIu} , \ \ \ \forall \mIu, \mIg \nonumber \\ 
 &  & &  \mPosRz{\mIg,\mIu} =\tilde{z}_{\mIg,\mIu} + \hat{g}_{\mIg,\mIu} S_z \mPosUyN{\mIu} + \delta_z  S_z \mPosUyN{\mIu}, \ \ \ \forall \mIu, \mIg \nonumber \\
 & & & -\frac{1}{2} \leq \delta_x   \leq  \frac{1}{2} , \ \ 
 -\frac{1}{2} \leq \delta_z   \leq  \frac{1}{2} \nonumber
 \end{align}
 which can be solved using a convex solver like CVXPY~{\cite{diamond2016cvxpy}.

 \paragraph{Complete Solution}
We have shown that for a given $\delta_x$ and $\delta_z$,  (\ref{prob:centralized_local}) gets simplified to  (\ref{prob:centralized_local_assignment}) which  can  be optimally solved. We also have shown that for a given $b_{\mIg,\mIu}$, we get  (\ref{prob:centralized_local_delta}) which can also be optimally solved. Hence, to solve (\ref{prob:centralized_local}), we use block coordinate descent. We optimize over each set of variables in an alternating manner, until the solution stops changing.  Since both  (\ref{prob:centralized_local_assignment}) and (\ref{prob:centralized_local_delta}) are solved to optimality, Problem (\ref{prob:centralized_local}) is guaranteed to converge to a stationary point~\cite[Prop. 2.7.1]{bertsekas_nonlinear_1999}. After solving, we obtain the optimal $\hat{\delta}_x$, $\hat{\delta}_z$, and $\hat{b}_{\mIg,\mIu}$ for all $\mIg$ and $\mIu$, along with $\hat{f}_{\mIg,\mIu}$ and $\hat{g}_{\mIg,\mIu}$. We need to substitute back to get the UAV positions. From   $\hat{b}_{\mIg,\mIu}$, the index of the placement assigned to the $\mIu$-th UAV  $\hat{\mIg}_{\mIu}$ is given by $\hat{\mIg}_{\mIu} = \underset{\mIg}{\text{argmax}}\hat{b}_{\mIg,\mIu}$. The assigned position is then calculated using
\begin{equation}
\mPosUx{\mIu} = \tilde{x}_{\hat{\mIg}_{\mIu},\mIu} + \hat{f}_{\hat{\mIg}_{\mIu},\mIu} S_x \mPosUyN{\mIu}  + \hat{\delta}_x  S_x \mPosUyN{\mIu} + \mPosUxB{\mIu}
\label{eq:x_subs}
\end{equation}
similarly for the z position
\begin{equation}
\mPosUz{\mIu} = \tilde{z}_{\hat{\mIg}_{\mIu},\mIu} + \hat{g}_{\hat{\mIg}_{\mIu},\mIu} S_z \mPosUyN{\mIu}  + \hat{\delta}_z  S_z \mPosUyN{\mIu} + \mPosUzB{\mIu}
\label{eq:z_subs}
\end{equation}
The centralized solution algorithm is summarized in Algorithm~\ref{alg:cent}. Hence, using block coordinate descent, we  obtained a suboptimal solution of~(\ref{prob:centralized_local}), which is a relaxation of~(\ref{prob:main}).  

\subsection{Upper Bound and Time Complexity}
The upper bound for the translation of UAVs is derived in Proposition~\ref{prop:uav_ub}
\begin{prop}
	\label{prop:uav_ub}
	The maximum absolute translation of  UAV $\mIu$  is upper bounded by $\frac{\sqrt{S^2_x+S^2_z}}{2}\mPosUyN{\mIu}$.
\end{prop}
\begin{IEEEproof}
See Appendix~C.
\end{IEEEproof}

 As for the algorithm computational complexity, it is the sum of the solution complexities of solving ({\ref{prob:centralized_local_assignment}}) and ({\ref{prob:centralized_local_delta}}) times the number of iterations. For the number of iterations, convergence typically occurred within fewer than five iterations, which can be enforced as a maximum number of  iterations.   The Hungarian algorithm used to solve  ({\ref{prob:centralized_local_assignment}})   has  complexity $\mathcal{O}(\mNg^3)$~{\cite{edmonds_theoretical_1972}}. For problem~({\ref{prob:centralized_local_delta}}), CVXPY~{\cite{diamond2016cvxpy}} uses   ECOS  second-order cone programming solver~{\cite{domahidi_ecos_2013}}, which relies on an interior-point algorithm based on Mehrorta predictor-corrector method. In general, the interior points algorithms' complexity depends on the number of variables~{\cite{potra_interior-point_2000}}. Since problem~({\ref{prob:centralized_local_delta}}) has only two variables ($\delta_x$ and $\delta_z$) regardless of the problem size, the solution time is dominated by the Hungarian algorithm. By  limiting the iterations to five, the complexity of Algorithm~{\ref{alg:cent}} is approximately $\mathcal{O}(\mNg^3)$. 

\begin{algorithm}[t!]
	\SetAlgoLined
	\SetKwInOut{Input}{input}
	\SetKwInOut{Output}{output}
	\Input{The initial positions of the UAV swarm $\{\mPosUI{0},\mPosUI{1},\cdots,\mPosUI{\mNu-1}\}$, The parameters of the GS $\mNgx,\mNgz,\mDgx,\mDgz$. The wavelength $\lambda$. }
	\Output{The optimized UAV positions.}
	current\_obj = $\infty$\;
	previous\_obj = 0\;
	Initialize $\delta_x=0$ and $\delta_z=0$ \;
	\While{\upshape {current\_obj-previous\_obj}>1e-5}{
		previous\_obj = current\_obj \;
		Solve (\ref{prob:centralized_local_assignment}) for $\{b_{\mIg,\mIu}\}$ using $\delta_x$ and $\delta_z$\;
		Solve (\ref{prob:centralized_local_delta}) for $\delta_x$ and $\delta_z$ using $\{b_{\mIg,\mIu}\}$ and assign the objective value to current\_obj \;
	}
	Calculate the position of UAVs, using (\ref{eq:x_subs}) and (\ref{eq:z_subs})
\caption{Centralized Solution} \label{alg:cent}
\end{algorithm}

\newcommand{\mFEx}[1]{ e^{[x]}_{#1}}

\section{Distributed Online Solution}
\label{sec:FF}
\begin{figure}[t!]
	\centering
	\includegraphics[width=3.5in]{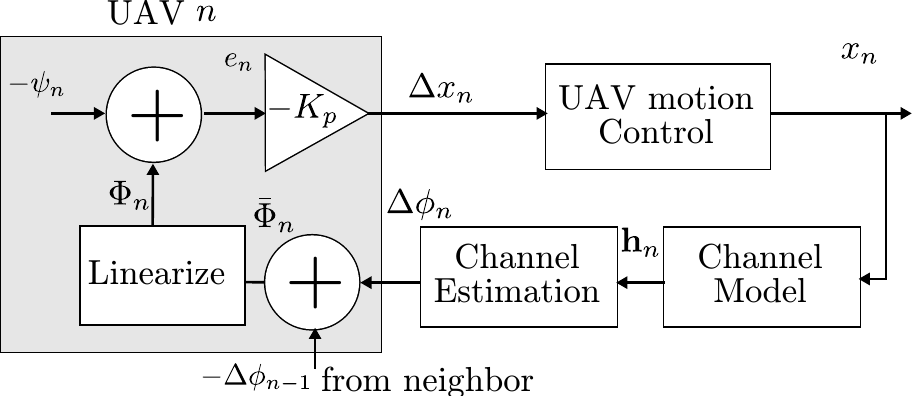}
	\caption{ The closed loop feedback system in UAV $\mIu$ where $\mIu \in \{ 1,\cdots,\mNu -1\}$. The calculations made on the UAV are highlighted in gray. The value $\Delta \phi_{\mIu-1}$ is obtained from the previous neighbor.}	
	\label{fig:force_field_control}
\end{figure}
In the case where the UAV positions are not known before deployment, we develop an iterative  distributed approach to be applied within the swarm in realtime. This approach uses channel estimates instead of positions for optimization and it attempts to minimize the inter-swarm communication overhead.
 In this approach, the UAVs agree on a formation, in which each UAV designates another UAV as its neighbor along each axis according to a criterion discussed later.  For a ULA GS, this formation is linear, and for a URA it is a rectangular grid. Each UAV estimates its channel and shares  the estimates with its neighbors using an ideal control side channel.  Using the neighbor channel measurements, the UAVs calculate an  error signal. This error signal drives a closed loop feedback  system, which decides the magnitude and direction of its motion. In this approach, each UAV moves based on information from its neighbor, as if each UAV exerts a force on its neighbor. Hence, we refer to this approach as Force Field (FF).  We start by deriving the fundamentals of this approach and show its convergence on a ULA GS aligned to the x-axis. Then, we discuss how it is applied to a URA GS. In the end, we discuss how the agreement on the formation is performed. 

\subsection{Fundamentals of Force Field}
The key idea behind FF is   that  for any optimal positions  $\mPmat \in \mPset$,  the equivalent $\mNg\times\mNg$ channel matrix $\mH$  can be shown to be a scaled and permuted  DFT matrix~\cite{haustein_ula_2003}.  The phase difference between successive elements of the $l$-th column of the DFT matrix is $\frac{2\pi}{\mNg}$. Hence, the phase difference at UAV $\mIu$ due to two consecutive ground antennas $l$ and $l+1$ is given by
\begin{equation}
	\Delta\phi_{\mIu} = \phi_{l,\mIu}- \phi_{l+1,\mIu} =  2\pi f_{\mIu}'+   \mIu  \left( \frac{2 \pi}{\mNgx} \right). \label{eq:cond_dft}
\end{equation}
where $\phi_{l,\mIu}=\angle \left[\mH\right]_{l,\mIu}$ is the phase of the channel between GS antenna $l$ and UAV $\mIu$ for some integer $f_{\mIu}'$. If (\ref{eq:cond_dft})  is satisfied for all UAVs $n\in\{0,\cdots\,\mNu-1\}$, and all GS antennas $l\in\{0,\cdots\,\mNg-2\}$,  the channel $\mH$ is an orthogonal scaled DFT matrix and the capacity is maximized. However, (\ref{eq:cond_dft}) determines the position of each UAV solely on its index $\mIu$ regardless of the remaining UAVs. This might lead to larger distance traveled since each UAV does not consider its neighbors' positions or channels.  Additionally, if a UAV   suffered from an external disturbance like wind, the remaining UAVs will not adapt.  To make each UAV adapt to its neighbors, we reformulate (\ref{eq:cond_dft})  to
\begin{equation}
 \Delta\phi_{\mIu}  -\Delta\phi_{\mIu-1} =    2\pi f_{\mIu} + \left( \frac{2 \pi}{\mNgx} \right) \label{eq:cond_ddft}
\end{equation}
for any integer $f_{\mIu}$. Here  we assume that the UAV formation has been established and UAV $\mIu-1$ shares its channel estimates with its neighbor UAV $\mIu$.   If all UAVs realize (\ref{eq:cond_ddft}), $\mH$ becomes orthogonal.  We define the difference between the right and the left sides of (\ref{eq:cond_ddft}) as  an error signal as follows 
 \begin{equation}
 \label{eq:ff_error}
e_{\mIu} = \Delta\phi_{\mIu}  -\Delta\phi_{\mIu-1} - \psi_{\mIu} 
 \end{equation}
where $  \psi_{\mIu}  = 2\pi f_{\mIu} + \left( \frac{2 \pi}{\mNgx} \right)$ is the target phase difference for some integer $f_{\mIu}$.    The objective of each UAV is to move such that this error signal is equal to zero.
We define $ \overline{\Phi}_{\mIu} = \Delta\phi_{\mIu}  -\Delta\phi_{\mIu-1}$ as the measured state of our system. When each UAV moves, this state changes and our goal is to make it equal to $\psi_{\mIu}$. Using  the distance approximation~(\ref{eq:dist_rel}) based on the far-field assumption, we get
 \begin{align}
 \overline{\Phi}_{\mIu}&= (\Delta\phi_{\mIu} - \Delta\phi_{\mIu-1})\%(2 \pi)  \label{eq:ff_meas}\\
 &\approx  \left( 2\pi \left(\frac{\mPosUx{\mIu}}{\mPosUyN{\mIu}}-\frac{\mPosUx{\mIu-1}}{\mPosUyN{\mIu-1}} \right) \frac{1}{S_x} \right) \%(2 \pi) \label{eq:ff_sys}
 \end{align}
 where the environment constant $S_x=\frac{\lambda \mR}{\mDgx}$ (as previously defined) and $\%$ denotes the modulus operator. The modulus operation accounts for phase wraps.  This relation would have been linear if it was not for the phase ambiguity. Using phase measurements only, we can not tell whether the UAV did not change position or moved to create a $2\pi$ phase difference. However, this ambiguity can be mitigated and the phase unwrapped by limiting the translation that each UAV performs at each step as discussed later. 
 After unwrapping the phase, the system becomes a linear system. Based on the error signal, each UAV can change its position to orthogonalize the channel using a closed loop feedback system. 
 
 \subsection{Force Field Algorithm}
The proposed feedback system is run iteratively in all UAVs. In  iteration $k$, all the UAVs move simultaneously, except the first UAV which is used as an anchor and does not move. This approach is described as follows:
Each UAV estimates its channel and calculates $\Delta\phi_{\mIu}[k]$ at iteration $k$. It shares this value with its direct neighbor in the formation, so that UAV $\mIu$ knows $\Delta\phi_{\mIu-1}[k]$ from its neighbor. 
 Each UAV calculates  the measured state $\overline{\Phi}_{\mIu}[k]$ and estimates the unwrapped phase state $\Phi_{\mIu}$ using
\begin{equation}
\label{eq:ff_phase_correction}
	\Phi_{\mIu}[k]=
	\begin{cases}
	 \overline{\Phi}_{\mIu}[k]  +  2 \pi \left\lfloor \frac{\Phi_{\mIu}[k-1]}{2\pi} \right\rfloor &  c=0  \\
	 \overline{\Phi}_{\mIu}[k] + 2\pi +  2 \pi \left\lfloor \frac{\Phi_{\mIu}[k-1]}{2\pi} \right\rfloor & c= 1 \\
	 \overline{\Phi}_{\mIu}[k] - 2\pi +  2 \pi \left\lfloor \frac{\Phi_{\mIu}[k-1]}{2\pi} \right\rfloor & c= 2 \\
	\end{cases}
\end{equation}
where
\begin{multline}
	c=\text{argmin}\{ |\overline{\Phi}_{\mIu}[k] -\overline{\Phi}[k-1]   |, 
	\ifdefined \singleCol\else	\\	\fi
	 |\overline{\Phi}_{\mIu}[k] + 2\pi- \overline{\Phi}[k-1]| ,  |\overline{\Phi}_{\mIu}[k] - 2\pi -\overline{\Phi}[k-1]|\}
\end{multline}
It is easy to verify that  phase wrap errors will not occur as long the phase transition between  iterations is less than $\pi$.  
 After linearizing the state, each UAV  calculates an error $e_{\mIu}[k]$  using (\ref{eq:ff_error}). 
Based on this error signal, it changes its position such that
\begin{equation}
\label{eq:ff_motion}
	\mPosUx{\mIu}[k+1] = \mPosUx{\mIu}[k] - K_p \mFEx{\mIu} [k]
\end{equation}
 where  $K_p$ is a  constant creating a proportional controller. The first UAV in the formation having $\mIu=0$ is used as an anchor, i.e, it does not change positions.  The closed loop  feedback system at UAV $\mIu$ is shown in Fig.~\ref{fig:force_field_control}, with the calculations that run in the UAV highlighted in gray. The input to our approach for UAV $\mIu$ is its phase estimates along with those of UAV $\mIu-1$. The output is the motion given by  $\Delta x_{n}[k] =\mPosUx{\mIu}[k+1] - \mPosUx{\mIu}[k]$. After a predefined  time sufficient for calculations in all UAVs, the output $\Delta x_{n}[k]$ is fed to the UAV motion control system  which navigates the UAV. After the swarm settles,  these steps can be repeated  for a fixed number of iterations until  $|\Delta x_{n}[k]|$ becomes small for all UAVs.  Note that since the system was linearized, instead of the proportional controller in~({\ref{eq:ff_motion}}),  more sophisticated controllers like PID can  speed up the convergence~{\cite{golnaraghi_automatic_2017}}.
 
 We now discuss the convergence of this approach for a single UAV and then generalize to the entire swarm.
\begin{lem}
	\label{lem:ff}
	The error in UAV $\mIu$ is guaranteed to converge to zero given that its previous neighbor, UAV $\mIu-1$, is fixed, if  $0<K_p< \frac{\mPosUyN{\mIu} S_x}{4\pi}$ .
\end{lem}
\begin{IEEEproof}
	See Appendix~D.
\end{IEEEproof}
\begin{thm}
	\label{thm:ff_conv}
The error of all UAVs is guaranteed to converge to zero if $0<K_p<\frac{ \underset{\mIu}{\min}(\mPosUyN{\mIu}) S_x}{4\pi}$.
\end{thm}
\begin{IEEEproof}
	 UAV  $0$ acts as an anchor and  does not move. Hence, the error of   UAV $1$, according to Lemma \ref{lem:ff} is guaranteed to converge to zero if $0<K_p< \frac{\mPosUyN{0} S_x}{4\pi}$. Once, it converges according to Lemma \ref{lem:ff}  the error of UAV $2$ is also guaranteed to converge to zero  if $0<K_p< \frac{\mPosUyN{1} S_x}{4\pi}$.  Similarly, we can show that all $\mNu$ UAVs will converge if $0<K_p< \frac{\underset{\mIu}{\min}(\mPosUyN{\mIu}) S_x}{4\pi}$. 
\end{IEEEproof}
Note that although  the UAVs closer to the fixed UAV converge first, all the non-converged UAVs move simultaneously.
As a consequence of simultaneous motion,  oscillations might occur; a UAV might move in some direction in an iteration and in the other direction in the following iteration because its previous neighbor has moved. A smaller value of $K_p$ will reduce the magnitude of  oscillations.

 \subsection{Force Field URA Extension}
Next, we discuss the extension from the ULA GS in the x-direction to a URA in the x-z plane. For a URA, each UAV needs to meet the orthogonality criterion~(\ref{eq:cond_ddft})  in both the x and z directions. The condition along the x-axis is
\begin{equation}
 \Delta\phi^{[x]}_{\mIu}  -\Delta\phi^{[x]}_{\mIu-1} =  \psi^{[x]}_{\mIu} \label{eq:cond_ddft_x}
\end{equation}
where  the superscript $[x]$ is to denote x-direction, $\psi^{[x]}_{\mIu}$ is the x phase  objective,  and 
$
	\Delta\phi^{[x]}_{\mIu} = \phi_{\mIu,\mIgx \mNgz+ \mIgz}- \phi_{\mIu,(\mIgx+1) \mNgz+ \mIgz} 
$
where $\mIgx \mNgz+ \mIgz$ and $(\mIgx+1) \mNgz+ \mIgz$ are the indices of two  consecutive GS antennas along the x-direction. Similar definitions exist for the z-direction using phase calculated for two consecutive GS antennas along the z-direction: 
$
	\Delta\phi^{[z]}_{\mIu} = \phi_{\mIu,\mIgx \mNgz+ \mIgz}- \phi_{\mIu,\mIgx \mNgz+ \mIgz+1} 
$.
To realize (\ref{eq:cond_ddft_x}) and its z equivalent,  FF is extended to apply the same procedures for a ULA along  both directions. Hence, each UAV needs to designate two neighbors, one for each direction. This makes the final FF formation a grid. This grid consists of $\mNgx$ lines applying linear FF along z direction and $\mNgz$ lines applying it along the x-direction. The anchor node that does not move in that case is a corner node having both grid indices $\mIux=\mIuz=0$. We note that for a ULA in the x-direction the set $\mPset$ is unconstrained in the z-direction.  Unlike the ULA, for the URA case, to retain orthogonality over the entire swarm, the UAVs  that form a line in the x-direction, need to  have the same phase with respect to the z-direction and vice versa. To accomplish that a small modification is made; the first line of the grid in the x-direction (having indices satisfying $\mIu\%\mNgz=0$ where $\%$ denotes the modulus operator) applies FF along the z-direction to have a phase difference along z equal to zero. The phase objective along the z direction $\psi^{[z]}_{\mIu}$ for UAV $\mIu$ in (\ref{eq:ff_error}) realizing this condition is
\begin{equation}
\label{eq:ff_phase_obj}
	\psi^{[z]}_{\mIu} = \begin{cases}
	0 &  \mIu\%\mNgz=0 \\
	2\pi/\mNgz & \text{otherwise}
	\end{cases}
\end{equation}
A similar relation can be derived for the phase objective along x. Since, for a URA, the same FF feedback system is applied along multiple lines with a minor modification, the  same convergence proofs apply.

 \subsection{Initializing Formations}
Last, we describe how the formations are established. As shown in Theorem 2, the convergence only depends on each UAV picking a node as a neighbor along each axis, such that all the UAVs create a grid formation.
The method of choosing the neighbor does not affect whether or not convergence will occur, however, it affects the distance that each UAV will travel to orthogonalize the channel. Our proposed approach relies on UAVs creating the formations based on an initial channel estimate that is shared globally among the swarm.

   After sharing the channel, all UAVs pick their closest neighbor based on phase relative to the  x-direction and then relative to the z direction.  Given that the measured phase states along x and z directions are defined as  $\overline{\Phi}^{[x]}_{\mNu} = \Delta\phi^{[x]}_{\mIu}  -\Delta\phi^{[x]}_{\mIu-1}$ and $\overline{\Phi}^{[z]}_{\mNu} = \Delta\phi^{[z]}_{\mIu}  -\Delta\phi^{[z]}_{\mIu-1}$ respectively. The assignment is accomplished in two stages, first, we sort the state along x such that  $\overline{\Phi}^{[x]}_{0} \leq \overline{\Phi}^{[x]}_{1} \leq \cdots \leq  \overline{\Phi}^{[x]}_{\mNu-1}$. Then,  each $\mNgz$ UAVs are divided into a group and sorted such that the  $m$-th group satisfies  $\overline{\Phi}^{[z]}_{m \mNgz}  \leq \overline{\Phi}^{[z]}_{m \mNgz +1}  \leq  \cdots \leq \overline{\Phi}^{[z]}_{ (m+1)\mNgz-1}$.
   This assignment guarantees that the phase along any line  of  UAVs in the grid is increasing.

  The entire force field algorithm is summarized in Algorithm~\ref{alg:ff}. The \textbf{forall} construct is used to indicate that all UAVs act in parallel. For simplicity, we consider using a fixed number of iterations $K_c$.  More adaptive stopping criteria can easily be developed based on the value of the error or the SINR. Since at convergence the interference among data streams is eliminated, the MIMO SINR is equal to the SNR when a single UAV is communicating with the GS. By setting the target SINR below the SNR, we can sacrifice the achievable capacity in favor of less  distance traveled by the UAVs. Next, we find an upper bound for the distance traveled. 
   \begin{prop}
   	\label{prop:ff_ub}
   	The distance traveled by UAV $\mIu$ when using Force Field is upper bounded by $\left(\sqrt{S^2_x+S^2_z}\right) (\max\{\mPosUyN{0},\mPosUyN{\mIu}\}) $  
   \end{prop}
   \begin{IEEEproof}
   	For the $m$-th line in the grid formed by the UAVs along the $x$ direction, UAVs are ordered such that 
   	\begin{equation}
   		-\pi \leq \overline{\Phi}^{[x]}_{m} \leq \overline{\Phi}^{[x]}_{ \mNgz+m} \leq \overline{\Phi}^{[x]}_{2 \mNgz+m} \leq \cdots \leq  \overline{\Phi}^{[x]}_{ (\mNgx-1) \mNgz + m}\leq \pi
   	\end{equation}
   	The first UAV is used as an anchor and it does not move. Each UAV is pushing its neighbor  to realize a phase difference of $\frac{2\pi}{\mNgx}$. Since, the formation guarantees that the UAVs are increasing in phase,  the worst-case scenario is when all UAVs start at exactly at the same phase.  In that case, the UAV having index $\mIu$ will have to travel  to create a phase difference of $\mIu \frac{2\pi}{\mNgx}$ from the start UAV.
   	Using (\ref{eq:ff_sys}), this is equivalent to having 
   	$
   		\left(\frac{\mPosUx{\mIu}}{\mPosUyN{\mIu}}-\frac{\mPosUx{0}}{\mPosUyN{0}} \right)=S_x
   	$. From which, $| \mPosUx{\mIu} -\mPosUx{0}| \leq S_x (\max\{\mPosUyN{0},\mPosUyN{\mIu}\}) $. A similar argument can be made for the z-direction. Combining both constraints, we get that the distance traveled by UAV $\mIu$ is upper bounded by $\left(\sqrt{S^2_x+S^2_z}\right) (\max\{\mPosUyN{0},\mPosUyN{\mIu}\})$. 
   \end{IEEEproof}
   We notice that the  traveled distance upper-bound for Force Field is higher than  the centralized solution upper-bound from Proposition~\ref{prop:uav_ub}. We also expect that the centralized approach would require less distance traveled than FF for several reasons; First, the centralized approach assumes the knowledge of the UAV initial positions, which define the problem. On the other hand, FF only uses only channel information, from which the positions can not be recovered. Second,   compared to the centralized solution, FF does not optimize the displacement of the entire swarm (from Lemma~\ref{lem:grid_shifts}) and just uses the first UAV as an  anchor. Having a fixed UAV  is crucial to guarantee the convergence as shown in Theorem~\ref{thm:ff_conv}.  Third, FF assigns the UAVs to the  positions in a simple way based on sorting the phases to avoid running a complicated assignment procedure in all the UAVs.  
   
   \begin{algorithm}[t!]
   	\SetAlgoLined
   	\SetKwInOut{Input}{input}
   	\SetKwInOut{Output}{output}
   	\Input{$\mNgx,\mNgz$}
   	\Output{Swarm positioned to maximize capacity.}
   	All UAVs estimates  channels and share it \;
   	\ForAll{\upshape UAV $n$ = 0 to $\mNu-1$}{
   		Sort phase estimates to identify neighbors\;
   		Calculate phase objective using (\ref{eq:ff_phase_obj}) along x and z\;
   	}
   	\For{\upshape iterations k = 1 to $K_c$}{
   		\ForAll{\upshape UAV $n$ = 0 to $\mNu-1$}{
   			Esitmate channel and share with  neighbors\;
   			Calculate  state (\ref{eq:ff_meas}) and linearize  (\ref{eq:ff_phase_correction}) in x and z\;
   			Calculate error using (\ref{eq:ff_error}) along x and z \;
   			Wait sufficiently for other UAVs calculations\;
   			Move in x and z according to (\ref{eq:ff_motion})\;
   			Wait sufficiently for other UAVs to move\;
   		}
   	}
   	\caption{Force Field Algorithm} \label{alg:ff}
   \end{algorithm}
 \subsection{Time Complexity  of Force Field}
Since FF is  a distributed algorithm, we discuss the complexity from the perspective of one UAV. In the initialization stage, each UAV has $2\mNg$ phase measurements from the entire swarm along x and z directions. Each UAV sorts the phases along x across all UAV and along z as groups of size $\mNgz$. Assuming the merge sort algorithm is used, the initialization  complexity is given by $\mathcal{O}(\mNg \log \mNg  + \mNgx \mNgz \log \mNgz)$. After  initialization, each UAV interacts only with one neighbor in the x-direction and one neighbor in the z-direction, regardless of the swarm size making the complexity be a function of only the number of iterations  $\mathcal{O}(K_c)$.  The fact that beyond  initialization FF complexity is independent of the swarm size makes it  scalable.
 
 \subsection{Comparison with Existing Distributed Algorithms }
 We briefly compare FF to Gradient Descent (GD) and Brute Force (BF) which were both proposed in~\cite{hanna_icassp_2019}. GD and BF are both iterative algorithms inspired by numerical optimization algorithm; gradient descent, and steepest descent respectively. GD relies on knowledge of the UAV positions and global channel knowledge within the swarm to calculate the gradient of the capacity with respect to positions. In each iteration,  in a sequential manner, all UAVs estimate the channel  and one UAV moves in the gradient descent direction.  BF also relies on global channel knowledge. In a BF iteration, a UAV takes 6 steps  in each of the 6 orthogonal directions. For each direction, the channel is estimated and the orthogonality of the channel is evaluated.  The UAV retains the position that improved the channel orthogonality. No upper bounds (UB) on distance traveled  nor convergence guarantees were derived for BG and GD in~\cite{hanna_icassp_2019}. Since GD and BF are based on numerical methods applied to a non-convex objective, it is not easy to analyze their convergence.
   Unlike BF and GD, in a FF iteration, each UAV only requires  knowledge of the channel from its direct neighbors reducing inter-swarm communications overhead. Also, in an FF iteration, all UAVs move simultaneously, thus requiring fewer channel estimates. The comparison between the algorithms is summarized in Table~\ref{tbl:comp}.
\begin{table}[t]
	\renewcommand{\arraystretch}{1.5}
	\caption{Distributed Algorithms Comparison \label{tbl:comp}}
	\centering
	\begin{tabular}{|c|c|c|c|}
		\hline
		Aspect & Force Field & Gradient Descent & Brute Force \\	\hline
		Channel Estimations & $K_c$ & $\mNu K_c$ & $6 \mNu K_c$\\ 	\hline 
		Inter-swarm Comm. & Neighbors & Swarm & Swarm\\	\hline 
		Convergence Proof & Yes & No & No\\	\hline
		Distance Upper Bound & Yes & No & No\\	\hline
	\end{tabular} 
\end{table}

\newcommand{\mW}{\b{W}}
\newcommand{\mV}{\b{V}}

\section{Optimizing a Linear Uplink Scenario}
\label{sec:linear_scenario}
So far we discussed algorithms that optimize UAV positions to maximize the channel capacity, which are applicable in the uplink and downlink scenarios regardless of the transmitter and receiver processing. Now, for an uplink scenario with UAVs as transmitters, we consider the joint optimization of the channel $\mH$ (through the UAV positions) and the linear precoders $\mV\in \C{\mNu \times \mNu}$ and combiners $\mW \in \C{\mNg \times \mNg}$. We can define the following joint optimization problem 
\begin{equation}
\underset{ \mH,\mW,\mV}
{\text{maximize}} \left\{ \log \det \left( \mI + \frac{1}{N_0 N_f} \left(\mW^H \mH \mV \right)^H \mW^H \mH  \mV\right)\right\}  
\label{eq:lin_prob}
\end{equation}  
where $N_0$ is the noise power spectral density and $N_f$ is the receiver noise figure, given that $\mH$ is a channel matrix defined according to our system model.  Each UAV is assumed to be carrying a radio with maximum transmitted power $P_T$. Since each UAV transmits an independent data stream,  $\mV$ is constrained to be a diagonal matrix. The columns of the combining matrix $\mW$ are assumed to be normalized.

If we define the equivalent channel $\mH_{eq}=\mW^H \mH \mV$, according to the upper bound~{(\ref{eq:cap_ub})}, the maximum occurs when $\mH_{eq}$ is orthogonal for a given SNR. Maximizing the SNR for an orthogonal $\mH_{eq}$ solves~({\ref{eq:lin_prob}}). Using our proposed position optimization algorithms, the channel $\mH$ can be made orthogonal. For orthogonal $\mH$,  the matched filter  combiner given by $\mW=\frac{\mH}{\mFb{\mH}/\mNg}$ makes $\mW^H \mH$ a scaled identity matrix and hence orthogonal.   The optimal diagonal precoding, in this case, is to use the maximum power $\mV=\sqrt{P_T}\mI$ to maximize the SNR\footnote{Note  that for iterative algorithms, before convergence, $\mH$ is not orthogonal and the proposed precoders and combiners are not necessarily optimal}. Hence, $\mH_{eq}$ is an orthogonal matrix maximizing the SNR and thus solves the joint optimization. Thus, we have derived the optimal linear precoders and combiners for an uplink scenario.

\section{Numerical Evaluation}
In this section, we evaluate the performance of the proposed algorithms using numerical simulations. The capacity improvements of position optimization are first evaluated along with their robustness to randomness due to the channel and UAV motion. Then, the convergence of Force Fields is evaluated under ideal and practical conditions along with other UAV positioning algorithms. The distance traveled per UAV for different swarm positions is then considered and compared to the derived upper bounds. Lastly, we evaluate the impact of position optimization on the capacity as we move to the massive MIMO regime with $\mNg>>\mNu$.
\begin{figure}[t!]
	\centering
	\includegraphics{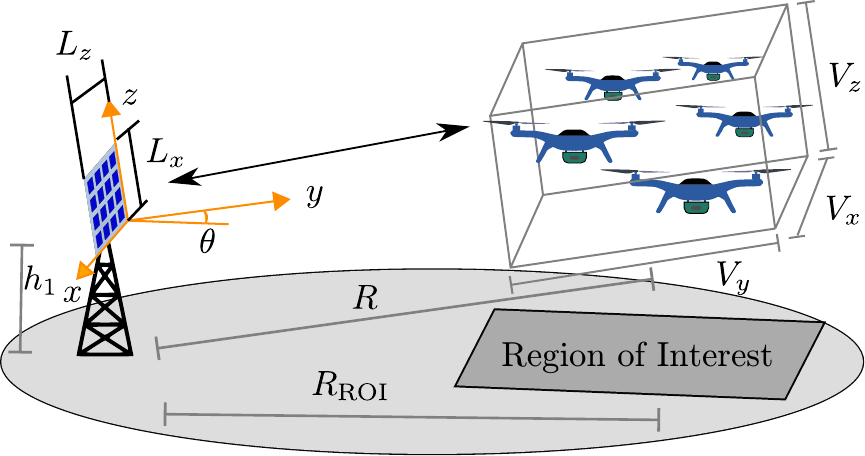}
	\caption{In the simulation setup, the UAVs are initialized in a rectangular area. The GS is tilted towards the swarm.}	
	\label{fig:simulation_model}
\end{figure}

\subsection{Simulation Setup}
We consider the simulation setup  shown in Fig.~\ref{fig:simulation_model}.   The GS consists of a URA  having aperture $L_x=\mDgx\mNgx=6$m and $L_z=\mDgz\mNgz=6$m operating at a frequency of 5GHz.  The large GS aperture reduces the distance traveled as shown in the derived upper bounds. However, an extremely large aperture is not practical.  Unless otherwise stated, we use  $\mNgx=6$ and $\mNgz = 2$ making  $\mDgx=1$m and $\mDgz=3$m.  The GS is  placed at a height $h_1=10m$, which is assumed to be higher than any surrounding buildings making a LOS path exist between the GS and swarm~\cite{chen_efficient_2020}. For UAVs deployed in a remote area, the GS can be adjusted to guarantee this condition. To account for non line-of-sight (NLOS) propagation paths, the channel is modeled as a Rician channel. The center of the region of interest is at a distance of $R_{\text{ROI}}=2$km from the UAV swarm.   For simplicity, the initial positions are randomly distributed in a rectangular parallelepiped having sides $V_x =10$, $V_y=300$, and $V_z=300$.   The elevation angle of the ground antenna used is $\theta=0.043$rad making the average height of the swarm approximately 100m.    %

We consider the uplink scenario, where the UAVs are the transmitters.    Channel estimation errors, when considered, are modeled using $\b{H}_\text{est}=\mH+\tilde{\b{H}}$, where $\b{H}_\text{est}$ is the estimated channel, and  $\tilde{\b{H}}$ is the   channel estimation error. The estimation error is modeled as a matrix with independent complex Gaussian elements with zero mean and  variance $\frac{1}{1+\text{SNR }T_\tau}$ where $T_\tau=10$ is the number of training symbols~\cite{hassibi_how_2003}. The UAV  motion errors, when considered, are modeled as an independent  random Gaussian  vector of size 3 having zero mean and a diagonal covariance matrix with a magnitude of 1m. This motion error vector is  added to the positions of the UAVs before channel estimation~\cite{hanna_icassp_2019}.  

  Each UAV has a transmit power $P_T=10$dBm and the  bandwidth used is assumed to be 1MHz~\cite{zeng_energy-efficient_2017}. The noise power spectral density used is $N_0=-174$dBm/Hz and the receiver has a noise figure $N_F$ of 3dB, making the noise power equal to -111 dBm.  
We use the sum rate obtained when using linear minimum mean square error (LMMSE) combiner at the GS as a metric~\cite[8.3.3]{tse_wireless_2005}. The LMMSE combining vector $w_\mIu$ for UAV $\mIu$ is calculated using $\b{w}_\mIu = \left(N_0 N_F \mI + \sum_{i=0,i\neq \mIu}^{\mNu} \mhc{i}^{[est]} \mhc{i}^{[est]H} \right)^{-1} \mhc{\mIu}^{[est]}$, where $\mhc{\mIu}^{[est]}$ is the $\mIu$-th column of the estimated channel. The signal-to-interference-and-noise ratio (SINR) of the $\mIu$-th stream is  given by
$  
\text{SINR}_\mIu = \frac{ P_T  |\b{w}_\mIu ^H\mhc{\mIu}^{[c]}|^2}{N_0 N_F +  P_T\sum_{i=0,i\neq \mIu}^{\mIu} |\b{w}_\mIu ^H \mhc{i}^{[c]}|^2}
$.
Using the SINR of each stream, the sum rate is calculated using $\text{SR}=\sum_{n=0}^{\mNu-1} \log{\left(1+\text{SINR}_\mIu\right)}$

For comparison, we consider relying on the randomness of the initial UAV positions  referred to as ``Init''. This is similar to what was proposed in~\cite{madhow_random_2013}, although we do not optimize the deployment region. We also consider positioning  the UAVs using the technique proposed for traditional planar uniform-rectangular arrays (URAs)~{\cite{bohagen_ura_2007}}.  We also consider BF and GD from~\cite{hanna_icassp_2019}. Although URAs were first proposed for fixed antenna arrays, they still can be used to maximize the capacity and along with uniform linear arrangements they have been proposed for UAVs~\mbox{\cite{su_maximum_2013,mozaffari_uav_array_2017,chandhar_massive_2018,hanna_spawc_2019}}.
\begin{figure}[t!]
	\centering
	\includegraphics{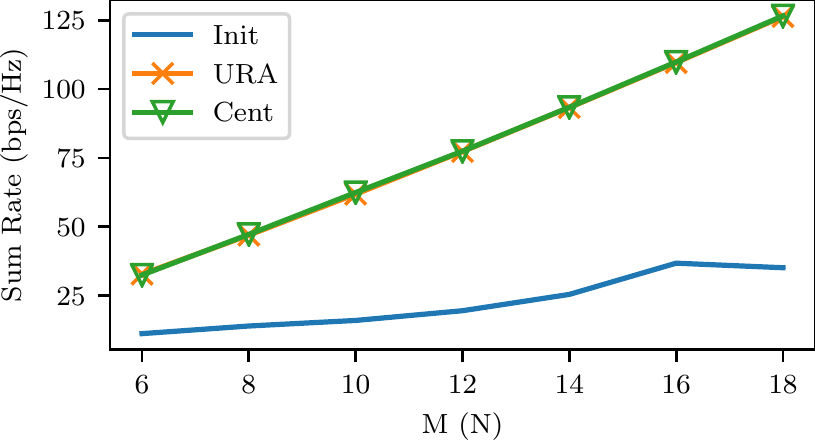}
	\caption{As we increase the number of UAVs and GS antennas, Cent gives an equal sum rate to URA, which is higher than relying on the initial positions.}	
	\label{fig:opt_rand_mmse}
\end{figure}
\subsection{Performance Gains of Position Optimization}
We start by demonstrating the performance gains that can be attained by optimizing the UAV swarm. We first  consider the case of the swarm and GS having an equal number of antennas $\mNg=\mNu$ and we vary $\mNgx$. This is shown in~Fig.~\ref{fig:opt_rand_mmse}.  As the $\mNgx$ increases, the sum rate of the optimized approaches (Cent and URA) increase linearly. This is expected from an optimized MIMO  channel.  However, this improvement comes at the cost of moving the UAVs from their initial positions. Unlike placing the UAV in a URA, our proposed approach minimizes the distance traveled. An example of a realization of random placement with $\mNgx=6$ ($\mNu=12$) is shown in Fig.~\ref{fig:opt_ura}. The initial placements of the UAVs are shown in blue and is assumed to be above the points of interest shown as crosses at $z=0$.  For URA shown in Fig.~\ref{fig:ura_3d},  UAVs need to travel  224m on average, which is far from the point of interest and might conflict with the objective of their deployment. On the other hand, for the centralized approach shown in Fig.~\ref{fig:opt_3d},   each UAV needs to travel only an average distance  of 20m from its initial position. This shows the limitation of relying on uniform placements.

\begin{figure}[t!]
	\centering
	\subfloat[Centralized.]{\includegraphics[scale=0.95]{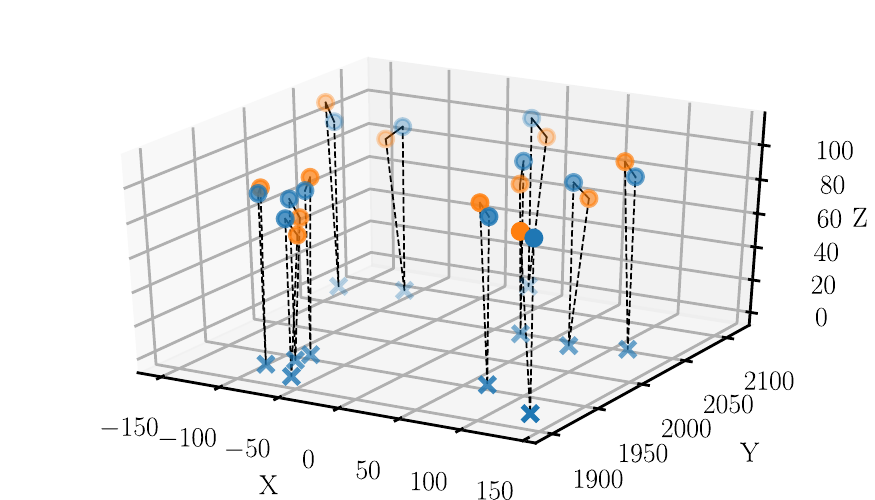}\label{fig:opt_3d}}
		\ifdefined \singleCol\else	\\	\fi
	\subfloat[URA.]{\includegraphics[scale=0.95]{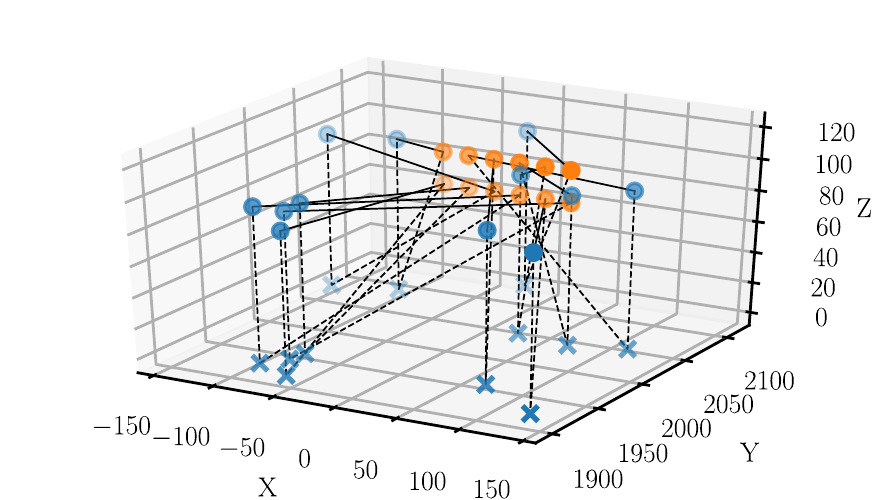}\label{fig:ura_3d}}
	\caption{Blue dots represent the initial positions, orange dots the final positions of the UAVs. The dashes on the ground represent the point of interest. While URA and Cent optimize capacity, URA moves the UAVs significantly far from their points of interest.}
	\label{fig:opt_ura}
\end{figure}

After attaining the centrally optimized positions, we evaluate the robustness of our obtained solution against external disturbances.  We consider the effects of NLOS propagation, channel estimation errors, and UAV motion errors. We vary the  value of the Rician K Factor,  and for each value, we simulate 100 random realizations of the Rician channel, localization errors, and channel estimation errors.  We plot the mean of the sum rates in Fig.~\ref{fig:opt_rand_rice_mmse} with the standard deviation  shown as error bars along with the single-user upper bound (UB) from (\ref{eq:cap_ub}). We see that for small values of the K-factor, the NLOS becomes dominant and both the optimized and non-optimized positions yield the same average capacity. As the LOS becomes more dominant and the K-factor increases, the optimized positions start approaching the capacity upper bound.  The random initial positions, on the other hand, converge to a lower sum rate. This is what we expect since an unoptimized LOS MIMO channel is correlated. In practice, the LOS air-to-ground channel typically has a high K-factor. In  channel measurement campaigns  performed at a frequency of 5-GHz (C-band)  for a LOS air-to-ground channel in near-urban and suburban environments, it was shown that the average K-factor was above 25dB~\mbox{\cite[Table V]{matolak_airground_2017}}.

\begin{figure}[t!]
	\centering
	\includegraphics{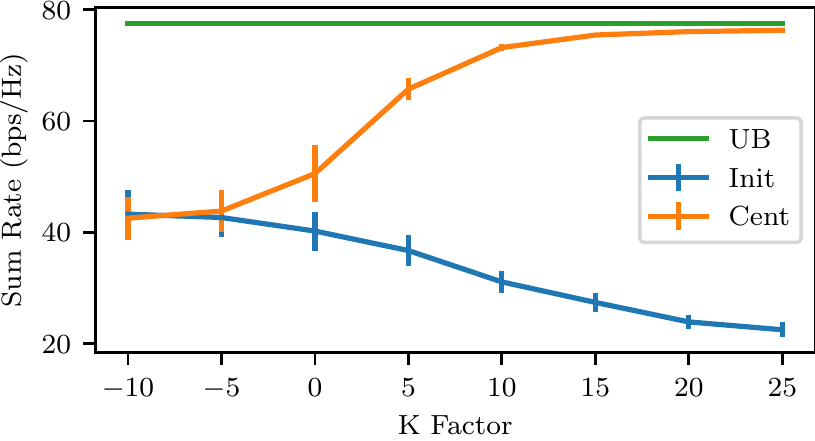}
	\caption{Optimizing positions improves capacity for LOS dominant channel having a large K-factor. Simulation is done for the placement in Fig.\ref{fig:opt_3d} and includes localization errors and channel estimation errors.}	
	\label{fig:opt_rand_rice_mmse}
\end{figure}

\subsection{Distributed Algorithms}

\begin{figure}[t!]
	\centering
	\subfloat[Average sum rate \label{fig:dist_alg_ideal_mmse}]{\includegraphics[scale=0.95]{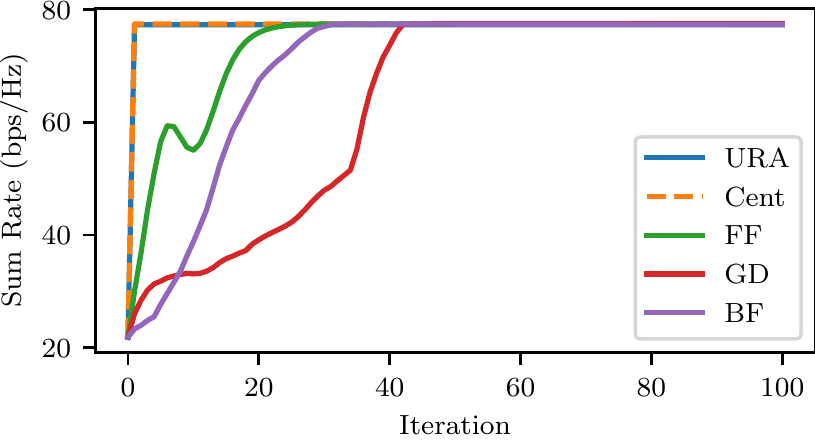}} \hspace{5mm} 
	\ifdefined \singleCol\else	\\	\fi
	\subfloat[Average distance traveled per UAV up to a given iteration. BF and URA  were omitted for   exceeding 100m. \label{fig:dist_alg_ideal_dist}]{\includegraphics[scale=0.95]{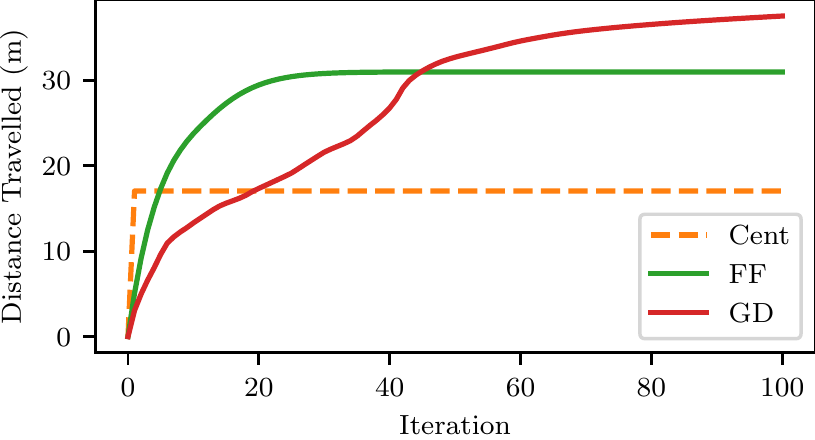}} 
	\caption{The sum rate and distance traveled for the ideal scenario.}	
	\label{fig:dist_alg_ideal}
\end{figure}

\begin{figure}[t!]
	\centering
	\includegraphics{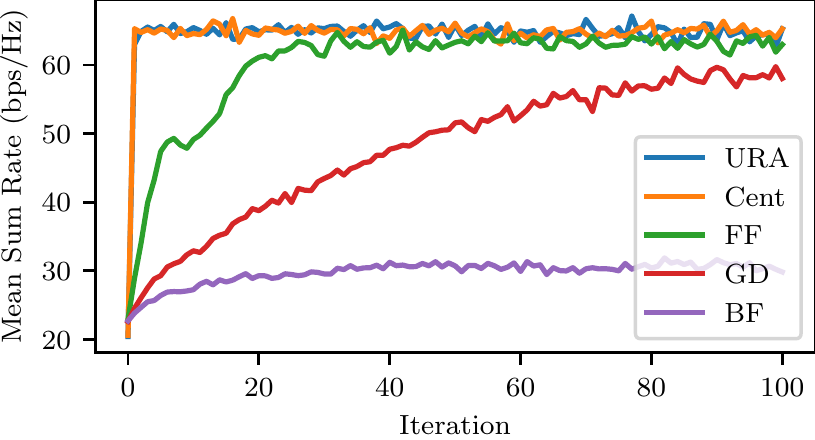}
	\caption{The sum rate of  distributed approaches under practical disturbances.}	
	\label{fig:dist_alg_mmse}
\end{figure}

Next, we evaluate the distributed algorithms performance in optimizing  positions. URA approach and the centralized algorithm (Cent) are used as benchmarks.  We compare  Force Field (FF) against  gradient descent (GD) and brute force (BF). For FF, we used $K_p=0.3\frac{\underset{\mIu}{\min}(\mPosUyN{\mIu}) S_x}{4\pi}$.
The convergence results   for 100 iterations in an ideal scenario with $K=\infty$  are shown in Fig.~\ref{fig:dist_alg_ideal}. From Fig.~\ref{fig:dist_alg_ideal_mmse}, we see that  all the methods converge to the optimal sum rate. The average distance traveled per UAV up to a given iteration is shown in Fig.~{\ref{fig:dist_alg_ideal_dist}} with the curves for URA and BF omitted for exceeding 100m. This Figure along with Fig.~{\ref{fig:dist_alg_ideal_mmse}} help characterize the tradeoff between attained capacity and the distance traveled.  We can see that after the first two iterations with only 12m average traveled distance, the attained capacity is doubled. This shows that, using FF, significant gains can be attained with a few  iterations and a limited  traveled distance.  Also  while moving to optimize the capacity,  the UAVs can work on their deployment tasks, hence FF does not impede on the deployment application.

  Then, we evaluate the convergence of these methods under practical disturbances. Namely, we consider a K-factor of 20dB along with channel estimation errors and localization errors added after each iteration. Also, in addition to the free space path loss, we consider log-normal  fading with  3.2dB standard deviation applied independently to each UAV~{\cite{matolak_airground_2017}}.  %
  Using the same initial positions, 100 realizations of these random distortions were simulated.   The average sum rate results are shown in Fig.~\ref{fig:dist_alg_mmse}. Compared to the ideal scenario, the sum rates even for URA and Cent are about 16\% lower because fading affects the channel magnitude and hence the SNR per stream. However,  we see that FF still converges to the sum rate bound attained by URA and Cent in about 30 iterations similar to the ideal scenario.  GD, on the hand, takes more than twice the iterations to converge compared the ideal scenario due to the random changes in the channel magnitude affecting the gradients. Brute Force  is severely impacted  by the motion errors and does not converge~\cite{hanna_icassp_2019}. Hence, our proposed FF is robust to practical disturbances expected in a swarm of UAVs and can attain the sum rate bound.   Compared to GD and BF, FF requires only a fraction of the inter-swarm communications and is guaranteed to converge within a bounded distance in an ideal scenario.

\subsection{Distance Traveled Per UAV}
UAV applications have different tolerance for UAV translations from the initial positions. Hence, it important to evaluate the distance that each UAV needs to travel.  To that end, we numerically evaluate the distance traveled by each UAV as a function of $\mR$. We consider 100 realizations, in which the UAVs are  initialized randomly in a cube such that $V_x=V_y=V_z=10$. The small cube guarantees that  the UAVs are within $\mFset$ as we change the distance $\mR$. In Fig.~\ref{fig:travel_ub}, the solid lines show the mean distance traveled  and the whiskers represent the range calculated over all realization and over the entire swarm. The upper bounds for Cent and FF derived in Propositions~\ref{prop:uav_ub} and~\ref{prop:ff_ub} respectively are plotted as dashed lines.  We can see that the distance that the UAVs need to travel increases as $\mR$ gets larger. This scaling is captured by our upper bounds, which can be used to estimate the worst case traveled distance. As expected,  the distributed algorithm using only channel estimates  requires a larger displacement than the centralized algorithm with perfect knowledge of the swarm initial positions. While we only show results for one center frequency and URA design, by using the upper bounds, verified in this section, we can predict the effect of changing the URA design  or the center frequency on the distance traveled.

\begin{figure}[t!]
	\centering
	\includegraphics{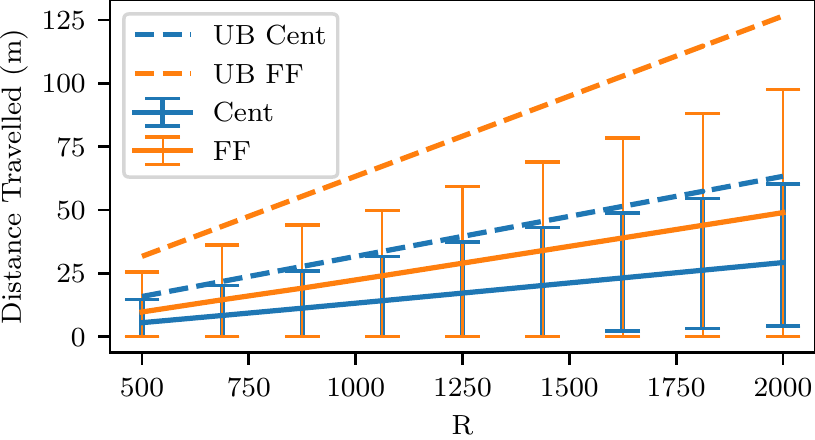}
	\caption{The mean distance traveled by the swarm. The error bars represent the range.}	
	\label{fig:travel_ub}
\end{figure}

\subsection{Massive MIMO Evaluation}
Now, we evaluate position optimization in the massive MIMO regime, where the number of GS antennas exceeds the number of UAVs $\mNg>>\mNu$.  Massive MIMO was shown to improve the capacity  by increasing the number of GS antennas~\cite{larsson_massive_2014}. One might presume that increasing the number of GS antennas eliminates the need for position optimization. We show that this is not the case.      We consider 8 UAVs and a GS with a fixed aperture such that $L_{x}=\mNgx\mDgx=4$ and $L_{z}=\mNgz\mDgz=6$. The ratio between antennas in the x and  z direction was set to be $\mNgx/\mNgz=2$ and the total number of GS antennas is increased~{\cite{martinez_experimental_2018}}.  From  Fig.~\ref{fig:eval_massive_mimo}, we see that the optimized approaches provide a higher sum rate than the non-optimized as expected. But as the number of antennas increases the sum rate gap between both optimized and non-optimized approaches does not converge to zero. The suboptimal massive MIMO performance in a LOS channel was also observed and analyzed in~\cite[Sec.~4.3]{ngo_aspects_2014}.   This indicates that even as the number of antennas increases, swarm optimization can provide significant improvements.  
 One way to interpret this result is to consider the grid of optimal positions similar to the one shown in Fig.~\ref{fig:optimal_set}. For a fixed aperture massive MIMO setup (assuming  the same x-z plane without loss of generality),  the smallest distance between two  optimal positions $\frac{S_x}{\mNgx}$  is constant and is equal to $\frac{\lambda \mR}{L_x}$.  This means that  by increasing the number of antennas, the optimal point density is the same. Hence, if the initial positions are far from any optimal ones, they will remain far as we increase $\mNg$. 
 
\begin{figure}[t!]
	\centering
	\includegraphics{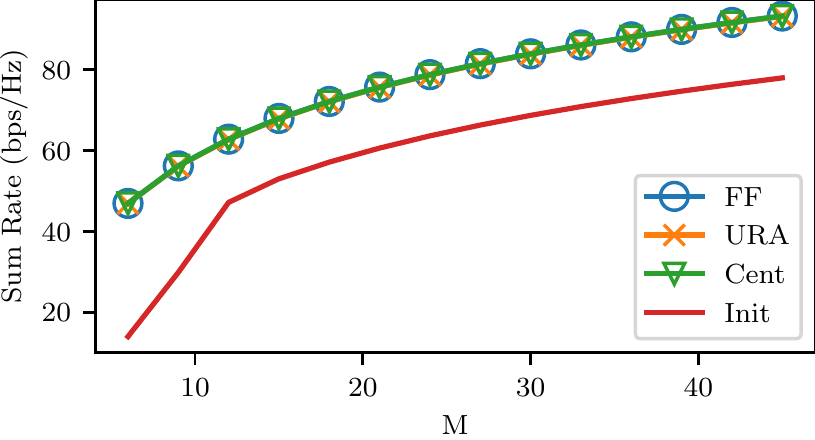}
	\caption{The sum rate of different approaches as we move to the massive MIMO regime.}	
	\label{fig:eval_massive_mimo}
\end{figure}

\section{Conclusion}
In this work, we  optimized the placements of a UAV swarm to maximize the MIMO backhaul capacity starting from given swarm  initial positions. We mathematically defined a set of UAV placements that orthogonalize the channel and maximize the MIMO capacity. The problem of minimizing the distance traveled to reach a placement in this set was formulated. An offline centralized solution was developed by  relaxing the problem and decomposing it into two convex problems which were solved iteratively using block coordinate descent. We also proposed FF as a distributed iterative solution to this problem. FF requires sharing channel estimates only between neighbors and we derived the conditions for its convergence. Using numerical simulation, we have shown its robustness under channel and UAV induced disturbances.  Upper bounds for the distance that  UAVs need to travel using the centralized solution and force field were derived and numerically verified. Our approaches were shown to provide significant sum rate improvements while requiring only bounded displacements. The gains from our approach were shown to remain significant as we transition to the massive MIMO regime with far more ground station antennas than UAVs.

\appendix
\subsection{Proof of Lemma~{\ref{lem:grid_rect}}}}
\label{ap:lema1}
	Let us define the scaled x and z translations, $\mPosRxd{\mIu} =   \frac{\mPosUx{\mIu}}{\mPosUy{\mIu}}$  and $\mPosRzd{\mIu} = \frac{\mPosUz{\mIu}}{\mPosUy{\mIu}}$. We start by assuming that the solution is found on a uniform grid with respect of the scaled variables with dimensions $\mNgx$ and $\mNgz$.   The separation of the UAVs along this grid in the x and z planes  is given by $\mDux$ and $\mDuz$, such that we can rewrite $\mPosRxd{\mIu} =\mIuxc \mDux $ and $\mPosRzd{\mIu} = \mIuzc\mDux $ for some integers $\mIuxc \in\{0,\cdots,\mNgx-1\} , \mIuzc \in\{0,\cdots,\mNgz-1\}$.  Our objective, hence, becomes calculating the value of $\mDux$ and $\mDuz$. Starting from (\ref{eq:orthogonal_column_approx}), we get
	
	\begin{align}
	\h{l}^{H} \h{k} & =  \sum_{\mIu=0}^{\mNu-1} \exp \left( 
	\frac{-j 2\pi}{\lambda} \left( (-\mIxl+\mIxk) \mDgx \mPosRxd{\mIu} + (-\mIzl+\mIzk) \mDgz \mPosRzd{\mIu} \right)
	\right) \\
	& = \sum_{\mIuxc=0}^{\mNgx-1} \sum_{\mIuzc=0}^{\mNgz-1} \exp \left( 
	\frac{-j 2\pi}{\lambda} \left( (-\mIxl+\mIxk) \mIuxc  \mDgx \mDux \right. \right.  \nonumber \\ 
	& \hspace{40mm}  \left. \left. + (-\mIzl+\mIzk) \mIuzc \mDgz  \mDuz \right)
	\right) \label{eq:breaking_sum}\\
	& = \sum_{\mIuxc=0}^{\mNgx-1} \exp \left( 
	\frac{-j 2\pi}{\lambda } \left( (\mIxk-\mIxl) \mIuxc  \mDgx \mDux \right) \right) \nonumber 
	\\& \hspace{10mm}  \cdot \sum_{\mIuzc=0}^{\mNgz-1} \exp \left( 
	\frac{-j 2\pi}{\lambda } \left( (\mIzk-\mIzl) \mIuzc \mDgz  \mDuz \right) \right) 
	\label{eq:two_geom}
	\end{align}
	In~(\ref{eq:breaking_sum}), the summation over UAVs was rewritten as a summation over the x and z UAV grid positions.
	As is evident from equation (\ref{eq:two_geom}), this summation is a product of two geometric sums and can therefore be simplified to 
	\begin{equation}
	\frac{\sin \left( \frac{\pi \mNgx (\mIxk-\mIxl)  \mDgx \mDux }{\lambda } \right) }{\sin\left(  \frac{\pi  (\mIxk-\mIxl)  \mDgx \mDux }{\lambda }   \right)} 
	\frac{\sin \left( \frac{\pi \mNgx (\mIzk-\mIzl)  \mDgz \mDuz }{\lambda } \right) }{\sin\left(  \frac{\pi  (\mIzk-\mIzl)  \mDgz \mDuz }{\lambda }   \right)} = 0
	\end{equation}
	where the summation is set to 0 because of the orthogonality condition defined in (\ref{eq:orthogonal_columns}). The orthogonality is achieved when $\mDux= \frac{\lambda}{\mNgx \mDgx}$ and $\mDuz= \frac{\lambda}{\mNgz\mDgz}$.
	If we set $\mPosUy{\mIu}$ to be constant for all UAVs, we get the same condition  derived for  the optimal  design of a parallel planar uniform rectangular arrays (URA) derived  in \cite{bohagen_ura_2007,larsson_ura_2005}. Hence, orthogonality is achieved when   $\mPosUx{\mIu} = \mIux \frac{\lambda \mPosUy{\mIu}}{\mNgx \mDgx}$ and $\mPosUz{\mIu} = \mIuz \frac{\lambda \mPosUy{\mIu}}{\mNgz \mDgz}$, where  $\mIu = \mIux \mNgz + \mIuz$.

\label{ap:prop1}
\subsection{Proof of Proposition~{\ref{prop:bounded_ints}}}
The objective of (\ref{eq:centr_prob1})  is monotonically increasing with respect to $(\mPosRx{\mIg,\mIu})^2$. This objective is minimized by minimizing $(\mPosRx{\mIg,\mIu})^2$. To prove that the optimal $f_{\mIg,\mIu}$ is within the set $\{-1,0\}$, we show that any value outside this set will correspond to a larger value of $(\mPosRx{\mIg,\mIu})^2$.

Given that $ 0 \leq \tilde{x}_{\mIg,\mIu} \leq  S_x \mPosUyN{\mIu} $ and that  $ -\frac{1}{2}   S_x \mPosUyN{\mIu} \leq \delta_x S_x \mPosUyN{\mIu} \leq\frac{1}{2}  S_x \mPosUyN{\mIu} $ from (\ref{eq:centr_prob1_delta}), we get
\begin{equation}
-\frac{1}{2}   S_x \mPosUyN{\mIu} \leq \tilde{x}_{\mIg,\mIu} + \delta_x S_x \mPosUyN{\mIu} \leq  \frac{3}{2} S_x \mPosUyN{\mIu} 
\end{equation}
So, for any value of  $\delta_x$ and $\tilde{x}_{\mIg,\mIu}$, using (\ref{eq:norm_x_trans})  the optimal value $\hat{f}_{\mIg,\mIu}$ can be calculated using 
\begin{equation}
\begin{aligned}
\hat{f}_{\mIg,\mIu}& =  \underset{f_{\mIg,\mIu}\in \Z}{\text{argmin}} ( \tilde{x}_{\mIg,\mIu} + f_{\mIg,\mIu} S_x \mPosUyN{\mIu}  + \delta_x  S_x \mPosUyN{\mIu})^2 \\
&=\begin{cases}
0 &  -\frac{1}{2}   S_x \mPosUyN{\mIu} \leq \tilde{x}_{\mIg,\mIu} + \delta_x S_x \mPosUyN{\mIu} <  \frac{1}{2} S_x \mPosUyN{\mIu} \\
-1 &  \frac{1}{2}   S_x \mPosUyN{\mIu} \leq \tilde{x}_{\mIg,\mIu} + \delta_x S_x \mPosUyN{\mIu} \leq  \frac{3}{2} S_x \mPosUyN{\mIu}  \\
\end{cases}%
\end{aligned}
\end{equation}
By substituting  $\hat{f}_{\mIg,\mIu}$ to calculate the  absolute translation $(\hat{x}'_{\mIg,\mIu})^2 = \underset{f_{\mIg,\mIu}\in \Z}{\text{min}} (\mPosRx{\mIg,\mIu})^2$, we find that it is  bounded by $(\hat{x}'_{\mIg,\mIu})\leq (\frac{S_x\mPosUyN{\mIu}}{2})^2$.	
If  $f_{\mIg,\mIu}$  is outside the set $\{-1,0\} $, we get a larger translation such that $(\mPosRx{\mIg,\mIu})^2 \geq \left(\frac{1}{2}   S_x \mPosUyN{\mIu}\right)^2$. Hence, the optimal value of $f_{m,n}$ has to be within $\{-1,0\}$ and is given by~(\ref{eq:fg_min}).
\subsection{Proof of Proposition~{\ref{prop:uav_ub}}}
\label{v}
In the proof of Proposition~\ref{prop:bounded_ints}, we showed  that  $(\hat{x}'_{\mIg,\mIu})^2\leq (\frac{S_x\mPosUyN{\mIu}}{2})^2$ and similarly $(\hat{z}'_{\mIg,\mIu})^2\leq (\frac{S_z\mPosUyN{\mIu}}{2})^2$. This holds for any  value of the remaining variables.
Hence, the translation made by UAV $\mIu$ is  upper bounded  by $\frac{\sqrt{S^2_x+S^2_z}}{2}\mPosUyN{\mIu}$.
\subsection{Proof of Lemma~{\ref{lem:ff}} }
	If $K_p$ is sufficiently small, the phase unwrapping given by (\ref{eq:ff_phase_correction}) retains the linearity of the measurements. Assuming  UAV $\mIu-1$ is fixed, $\mPosUx{\mIu-1}$ is constant across iterations, and we get
	\begin{equation}
	\begin{aligned}
	e_{\mIu}[k] &=  2\pi \left(\frac{\mPosUx{\mIu}[k]}{\mPosUyN{\mIu}}-\frac{\mPosUx{\mIu-1}}{\mPosUyN{\mIu-1}} \right) \frac{1}{S_x} -  \psi_{\mIu}  \\
	&= 2\pi \left(\frac{\mPosUx{\mIu}[k-1]-K_p e_{\mIu} [k-1]}{\mPosUyN{\mIu}}-\frac{\mPosUx{\mIu-1}}{\mPosUyN{\mIu-1}} \right) \frac{1}{S_x}  - \psi_{\mIu} \\
	&=  -2\pi\frac{K_p e_{\mIu} [k-1]}{\mPosUyN{\mIu} S_x} +2\pi \left(\frac{\mPosUx{\mIu}[k-1]}{\mPosUyN{\mIu}}-\frac{\mPosUx{\mIu-1}}{\mPosUyN{\mIu-1}} \right) \frac{1}{S_x} - \psi_{\mIu} \\
	& = \left(  1 -\frac{K_p 2\pi }{\mPosUyN{\mIu} S_x}\right) e_{\mIu} [k-1]
	\end{aligned}
	\end{equation}
	If $0<K_p< \frac{\mPosUyN{\mIu} S_x}{2\pi}$, the error will decrease in each iteration, and hence it will converge to zero. However, to avoid phase wrap errors when using (\ref{eq:ff_phase_correction}) we need to guarantee that any transition does not exceed $\pi$, which is realized when $0<K_p<\frac{\mPosUyN{\mIu} S_x}{4\pi}$

\ifCLASSOPTIONcaptionsoff
  \newpage
\fi

\bibliographystyle{IEEEtran}
\bibliography{references}

\begin{IEEEbiographynophoto}{Samer Hanna}
	received the B.Sc. degree in Electrical Engineering from Alexandria University, Alexandria, Egypt in 2013, and the M.Sc. degree in Engineering Mathematics from the same university in 2017. He is currently  pursuing  a  Ph.D. degree  at  the  University  of  California,  Los Angeles, CA, USA. His research interests include the applications of machine learning  in wireless communications and coordinated communications using  unmanned aerial vehicles .
\end{IEEEbiographynophoto}

\begin{IEEEbiographynophoto}{Enes Krijestorac}
	received a B.S. degree in Electrical Engineering from New York University, Abu Dhabi in 2018, graduating \textit{summa cum laude}. He is currently pursuing a Ph.D. degree at the University of California, Los Angeles, US. His research interests include UAV assisted wireless communication, modelling of wireless communication using machine learning and distributed computing systems. 
\end{IEEEbiographynophoto}

\begin{IEEEbiographynophoto}{Danijela Cabric}
is Professor in Electrical and Computer Engineering at University of California, Los Angeles. She earned MS degree in Electrical Engineering in 2001, UCLA and Ph.D. in Electrical Engineering in 2007, UC Berkeley, Dr. Cabric received the Samueli Fellowship in 2008, the Okawa Foundation Research Grant in 2009, Hellman Fellowship in 2012 and the
National Science Foundation Faculty Early Career Development (CAREER) Award in 2012 and Qualcomm Faculty award in 2020. She served as an Associate Editor of IEEE Transactions of Cognitive Communications and Networking, IEEE Transactions of Wireless Communications, IEEE Transactions on Mobile Computing and IEEE Signal Processing Magazine, and IEEE ComSoc Distinguished Lecturer. Her research interests are millimeter-wave communications, distributed communications and sensing for Internet of Things, and machine learning for wireless networks co-existence and security. She is an IEEE Fellow.
\end{IEEEbiographynophoto}

\end{document}